\documentclass[11pt,a4paper]{article}
\usepackage{jcappub,afterpage}

\title{Photon and neutrino redshift in the field of braneworld compact stars}

\author[a]{Jan Hlad\'{\i}k}
\author[a]{and Zden\v{e}k Stuchl\'{\i}k}

\affiliation[a]{Institute of Physics, Faculty of Philosophy and Science, Silesian University in Opava,\\ Bezru\v{c}ovo n\'{a}m. 13, CZ-746 01 Opava, Czech Republic}

\emailAdd{jan.hladik@fpf.slu.cz}
\emailAdd{zdenek.stuchlik@fpf.slu.cz}

\abstract{We study gravitational redshift of photons and neutrinos radiated by the braneworld neutron or quark stars that are considered in the framework of the simple model of the internal spacetime with uniform distribution of energy density, and the external spacetime described by the Reissner-Nordstr\"{o}m geometry characterized by the braneworld ``tidal'' charge $b$. For negative tidal charges, the external spacetime is of the black-hole type, while for positive tidal charges, the external spacetime can be of both black-hole and naked-singularity type. We consider also extremely compact stars allowing existence of trapped null geodesics in their interior. We assume radiation of photons from the surface at radius $R$, neutrinos from the whole compact star interior, and their motion along radial null geodesics of the spacetime. In dependency on the compact stars parameters $b$ and $R$, the photon surface redshift is related to the range of the neutrino internal redshift and the signatures of the tidal charge and possible existence of extremely compact stars are discussed. When both surface (photon) and internal (neutrino) redshift are given by observations, both compact star parameters $R$ and $b$ can be determined in the framework of our simple model.}

\keywords{modified gravity, neutrino astronomy, neutron stars}

\begin{document}
\maketitle
\flushbottom

  \section{Introduction}\label{SECintro}
    In the braneworld models, the observable universe is a 3-brane (domain wall) to which the matter fields are confined, while the gravity field enters the extra spatial dimensions of size that could strongly exceed the Planck length scale \cite{Ark-Dim-Dva:1998:}. Gravity can be localized near the brane at low energies even with an infinite size extra dimension with warped spacetime satisfying the 5D Einstein equations containing negative cosmological constant \cite{Ran-Sun:1999:} and an arbitrary energy-momentum tensor allowed on the brane \cite{Shi-Mae-Sas:1999:}. The Randall-Sundrum model gives 4D Einstein gravity in low energies, while significant deviations from the Einstein gravity occur at very high energies, in the early universe and in vicinity of compact objects as black holes and neutron stars \cite{Maar:2004:}.

    Although no exact solution of the full 5D Einstein equations is know recently, there is a variety of astrophysically plausible special solutions of the 4D effective Einstein equations constrained to the brane. Such solutions describe black holes with spherical symmetry \cite{Dad-Maar-Pap-Rez:2000:PHYSR4:} and axial symmetry \cite{Ali-Gum:2005:}, or compact objects that could represent neutron (quark) stars \cite{Ger-Maar:2001:}, but there is no guarantee that these solutions have a regular bulk spacetime. The black hole spacetimes are determined by geometry of the spherical Reissner-Nordstr\"{o}m (RN) and axial Kerr-Newman (KN) type where the electric charge squared is substituted by the braneworld ``tidal charge'' parameter representing the tidal (Weyl tensor) effects of the bulk space onto the 4D black hole structure. Astrophysically relevant properties of the braneworld black holes were studied in a series of papers devoted to both motion of matter in their vicinity \cite{Stu-Kot:2009:,Ali-Tal:2009:,Abdu-Ahme:2010:PHYSR4,Mam-Hak-Toj:2010:MPLA:,Mor-Ahme-Abdu-Mam:2010:ASS:} and optical phenomena \cite{Sche-Stu:2009:a,Sche-Stu:2009:b,Bin-Nun:2010:PHYSR4:a,Bin-Nun:2010:PHYSR4:b}.

    In the Randall-Sundrum model, full 5D bulk spacetime solutions for relativistic compact stars and black holes have to be inevitably solved by numerical methods by properly imposing regularity conditions in the bulk. For uniform energy density stars such a solution was obtained and discussed by Wiseman, who demonstrated possible existence of very compact stars~\cite{Wis:2002:}. Using the numerical methods developed in~\cite{Wis:2002:,Wis:2003:}, some 5D black hole solutions were constructed for limited radii~\cite{Kud-Tan-Nak:2003:,Kud:2004:}. Very recently, a black hole 5D solution localized to a brane has been numerically constructed and existence of dynamically stable black holes was demonstrated for all radii~\cite{Fig-Wis:2011:}. Here, we restrict our attention to the solution of the effective Einstein equations that can be given in a simple, analytic form.

    In the simple model of spherically symmetric stars with uniform energy density profile a variety of special solutions having asymptotically Schwarzschildian character and satisfying the braneworld boundary conditions were found for the effective Einstein equations~\cite{Ger-Maar:2001:}. The most popular is the one with external spacetime described by the Reissner-Nordstr\"{o}m geometry with the braneworld tidal charge parameter $b$ reflecting the tidal effects of the bulk and related to the energy density and brane tension. Usually, the braneworld tension is assumed positive while the related tidal charge has to be negative \cite{Ger-Maar:2001:}, but the negative tension and related positive tidal charge are not excluded \cite{Dad-Maar-Pap-Rez:2000:PHYSR4:}, so we consider here both positive and negative tidal charge of the compact object. Notice that in such a case, the exterior of a neutron star with positive charge can be described by both black-hole and naked-singularity types of the external RN spacetime (for details see \cite{Kot-Stu-Tor:2008:CLASQG:}). In the later case, it should be stressed that there is no naked singularity in the solution as this part of the solution is replaced by a star. Properties of the compact stars with RN external geometry were extensively studied both in the weak field limit \cite{Boh-Har-Lob:2008:CLAQG:,Boh-Ris-Har-Lob:2010:CLASQG} and strong field limit when some restrictions on the brane tension were implied from the data of kHz quasiperiodic oscillations (QPOs) observed in low mass X-ray binary systems containing a neutron (quark) star \cite{Kot-Stu-Tor:2008:CLASQG:} or for trapping of neutrinos in extremely compact stars allowing for existence of trapped null geodesics \cite{Stu-Hla-Urb:2011:}.

    Here we focus to the tidal charge influence on the redshift of photons radiated from the compact star surface and neutrinos radiated from the whole compact star interior. For simplicity, we assume the photon and neutrino motion along radial null geodesics that are usually considered in estimating the parameters of neutron (quark) stars~\cite{Lat-Pra:2007:PhysRep:}. The neutron star is considered in the framework of the simple model with external spacetime described by the RN geometry --- then our results can be compared to a wide variety of recent studies related to this kind of the braneworld solutions. We also consider Extremely Compact Stars (ECS) containing trapped null geodesics \cite{Abr-Mil-Stu:1993:PHYSR4:,Stu:2000:ACTPS2:,Nil-Cla:2000:GRRelStarsPolyEOS:,Stu-Tor-Hle-Urb:2009:}. For negative tidal charges, their external spacetime is of the black-hole type with one unstable photon circular orbit and the ECS surface at radius $R$ has to be located under the photon circular orbit. For positive tidal charges, their external spacetime can be of both black-hole (with an unstable photon circular orbit) and naked-singularity type when two photon circular orbits (stable and unstable) can exist, or none such orbit exists. In the case of positive tidal charges, the ECS can exist with surface radius $R$ located both under and above the unstable photon circular orbit (which is located above the stable photon circular orbit in the naked-singularity type external spacetimes) and can exist even with external naked-singularity spacetimes allowing none photon circular orbit \cite{Stu-Hla-Urb:2011:}

    Our paper is organized as follows. In Section~\ref{SECbrwns}, we summarize properties and matching conditions of the internal uniform energy density spacetime and the Reissner-Nordstr\"{o}m external spacetime. Null geodesics of both the internal and external spacetime are described in terms of properly given effective potential and the ECS are classified according to the properties of the trapping region of the null geodesics. In Section~\ref{SECredsh}, the surface redshift and the redshift profile through the internal spacetime are determined under assumption of purely radial motion, and their properties are discussed, especially with respect to the ECS and photon circular orbits of the external spacetime. In Section~\ref{SECconcl}, observational characteristic of the redshift profiles are introduced and possibilities to determine the surface radius and tidal charge are discussed. Finally, concluding remarks are presented. Throughout the paper, we shall use the high-energy units with $\hbar = c = k_\mathrm{B} = 1$, if not stated otherwise.

  \section{Braneworld neutron stars}\label{SECbrwns}
    We consider the braneworld model of neutron stars with spherical symmetry and uniform distribution of the energy density in their interior. The external spacetime is given by the Reissner-Nordstr\"{o}m geometry with the braneworld tidal charge representing the influence of the tidal effects of the bulk space \cite{Dad-Maar-Pap-Rez:2000:PHYSR4:}.

  \subsection{Internal and external spacetime and matching conditions}
    In the standard Schwarzschild coordinates and the high-energy units, the line element of the spherically symmetric spacetimes reads

    \begin{equation}
      \mathrm{d}s^2 = -A^{2}(r)\mathrm{d}t^2 + B^{2}(r)\mathrm{d}r^2 + r^{2}\mathrm{d}\Omega.
    \end{equation}

    The internal solution, matched to the external geometry at the surface of the star $r = R$, is characterized by the uniform energy density distribution --- $\varrho = \mathrm{const}$, and by the tension of the brane --- $\lambda$. We assume $\varrho > 0$ and both possibilities $\lambda > 0$, $\lambda < 0$ for the brane tension, although the positive value is more realistic \cite{Dad-Maar-Pap-Rez:2000:PHYSR4:}. The line element of the internal geometry is given by the metric coefficients $A^{-}(r)$, $B^{-}(r)$ that are determined by \cite{Ger-Maar:2001:}

    \begin{equation}
      A^{-}(r) = \frac{\Delta(R)}{\left(1 + p(r)/\varrho\right)}
    \end{equation}
    with the pressure radial profile given by
    \begin{equation}
      \frac{p(r)}{\varrho} = \frac{\left[\Delta(r) - \Delta(R)\right]\left(1 + \frac{\varrho}{\lambda}\right)}{3\Delta(R) - \Delta(r) + \left[3\Delta(R) - 2\Delta(r)\right]\left(\frac{\varrho}{\lambda}\right)}.
    \end{equation}
    and
    \begin{equation}
      \left(B^{-}(r)\right)^2 = \frac{1}{\Delta^{2}(r)} = \left[1 - \frac{2G M}{r}\left(\frac{r}{R}\right)^3 \left(1 + \frac{\varrho}{2\lambda}\right)\right]^{-1},
    \end{equation}
    where $M = \frac{4}{3}\pi\varrho R^3$. The maximum of the pressure profile is at $r = 0$, while there is $p(r)/\varrho = 0$ at $r = R$.

    The reality condition on the metric coefficient $B^{-}(r)$ (taken at $r = R$) implies a relation between $\lambda$, $\varrho$ and $R$ that can be expressed in the form
    \begin{equation}\label{EQRm2GM}
      \frac{G M}{R - 2 G M} \geq \frac{\varrho}{\lambda}.
    \end{equation}
    Considering the restriction $R > 2 G M$ ($R < 2 G M$), we can see that the reality condition~(\ref{EQRm2GM}) is satisfied for all $\lambda < 0$ (forbidden for all $\lambda > 0$), while for positive tension $\lambda > 0$ (negative tension $\lambda < 0$), we obtain a limit on the positive (negative) tension given by~\cite{Ger-Maar:2001:}
    \begin{equation}
      \lambda \geq \left(\frac{R - 2 G M}{G M}\right)\varrho .
    \end{equation}

    The line element of the external geometry is given by the metric coefficients $A^{+}(r)$, $B^{+}(r)$ that are determined by
    \begin{equation}
      \left( A^{+}(r)\right)^{2} = \left( B^{+}(r)\right)^{-2} = 1 - \frac{2G\mathcal{M}}{r} + \frac{q}{r^2},
    \end{equation}
    where, due to the matching conditions on the neutron star surface, i.e., $A^{-}(R) = A^{+}(R)$, $B^{-}(R) = B^{+}(R)$, the external mass parameter $\mathcal{M}$ and external tidal charge parameter $q$ are  related to the internal geometry parameters by $\varrho$, $\lambda$ (and $M$, $R$) by the relations \cite{Ger-Maar:2001:}

    \begin{equation}
      q = -3 G M R\frac{\varrho}{\lambda}
    \end{equation}
    \begin{equation}
      \mathcal{M} = M \left(1 - \frac{\varrho}{\lambda}\right).
    \end{equation}
    For $\lambda > 0$, the tidal charge $q < 0$ and $\mathcal{M} < M$, while for $\lambda < 0$, there is $q > 0$ and $\mathcal{M} > M$. Notice that for $\lambda > 0$, the condition $\varrho < \lambda$ has to be satisfied in order to have $\mathcal{M} > 0$.

    For our purposes, it is convenient to express the internal spacetime using the parameters of the external spacetime that can be directly determined from observations of accretion and optical phenomena in vicinity of the neutron (quark) stars. Since the matching conditions imply the relations
    \begin{equation}
      q = - \frac{3 G M R \varrho}{\lambda} = \frac{3 G \mathcal{M} R}{\left(1 - \frac{\lambda}{\varrho}\right)},
    \end{equation}

    \begin{equation}
      \frac{\lambda}{\varrho} = 1 - \frac{3 G \mathcal{M} R}{q},
    \end{equation}
    the internal metric can be expressed in terms of the gravitational radius $r_\mathrm{g}$ and dimensionless braneworld tidal charge $b$
    \begin{equation}
      r_\mathrm{g} \equiv G \mathcal{M},\qquad b \equiv \frac{q}{r_\mathrm{g}^{2}},
    \end{equation}
    in the form
    \begin{equation}
      A^{-} (r) = \frac{3\Delta(R)\left[\Delta(R) \left(2b - 3 R/r_\mathrm{g}\right) + \Delta(r)\left(R/r_\mathrm{g} - b\right)\right]}{2\Delta(R)\left(2b - 3R/r_\mathrm{g}\right)-\Delta(r)b},
    \end{equation}
    \begin{equation}
      \left(B^{-}(r)\right)^2 = \Delta^{-2}(r) = \left[1 - \frac{r_\mathrm{g}}{r}\left(\frac{r}{R}\right)^3 \left(2 - b\frac{r_\mathrm{g}}{R}\right)\right]^{-1},
    \end{equation}
    \begin{equation}
      \Delta^2 (R) = 1 - 2 \frac{r_\mathrm{g}}{R} + b \left(\frac{r_\mathrm{g}}{R}\right)^{2}.
    \end{equation}

    In the following, we shall use the dimensionless units ($r_\mathrm{g} = 1$). Then pressure function reads
    \begin{equation}
      \frac{p}{\varrho}\left(r, R, b\right) = \frac{\left(\Delta(r) - \Delta(R)\right) \left(2b - 3R\right)}{3 \Delta(R) \left(2b - 3R\right) - 3\Delta(r) \left(b - R\right)}.
    \end{equation}
    The pressure increases monotonously with radius decreasing and its central value is given by
    \begin{equation}
      \frac{p}{\varrho}\left(r = 0, R, b\right) = \frac{3R - 2b}{3R \left(\sqrt{b + (R - 2)R} + R - 3\right) + 6b}.
    \end{equation}

  \subsection{Limits on existence of uniform density stars}
    The limit on the existence of the uniform density spherical configuration is related to compactness of the star and is given by the condition of pressure finiteness in the center. The limit on compactness of the star expressed in terms of the gravitational mass related to the internal spacetime reads~\cite{Ger-Maar:2001:}
    \begin{equation}\label{EQGMR}
      \frac{G M}{R}\leq \frac{4}{9}\left[\frac{1 + \frac{5}{4}\frac{\varrho}{\lambda}}{\left(1 + \frac{\varrho}{\lambda}\right)^2}\right].
    \end{equation}
    In terms of the external gravitational mass parameter $\mathcal{M}$, the compactness limit~(\ref{EQGMR}) is transformed to the form
    \begin{equation}
      \frac{G \mathcal{M}}{R}\leq \frac{4}{9}\left[\frac{\left(1 + \frac{5}{4}\frac{\varrho}{\lambda}\right)\left(1 - \frac{\varrho}{\lambda}\right)}{\left(1 + \frac{\varrho}{\lambda}\right)^2}\right].
    \end{equation}

    \begin{figure}[t]
      \centering\includegraphics[width=0.8\hsize,keepaspectratio=true]{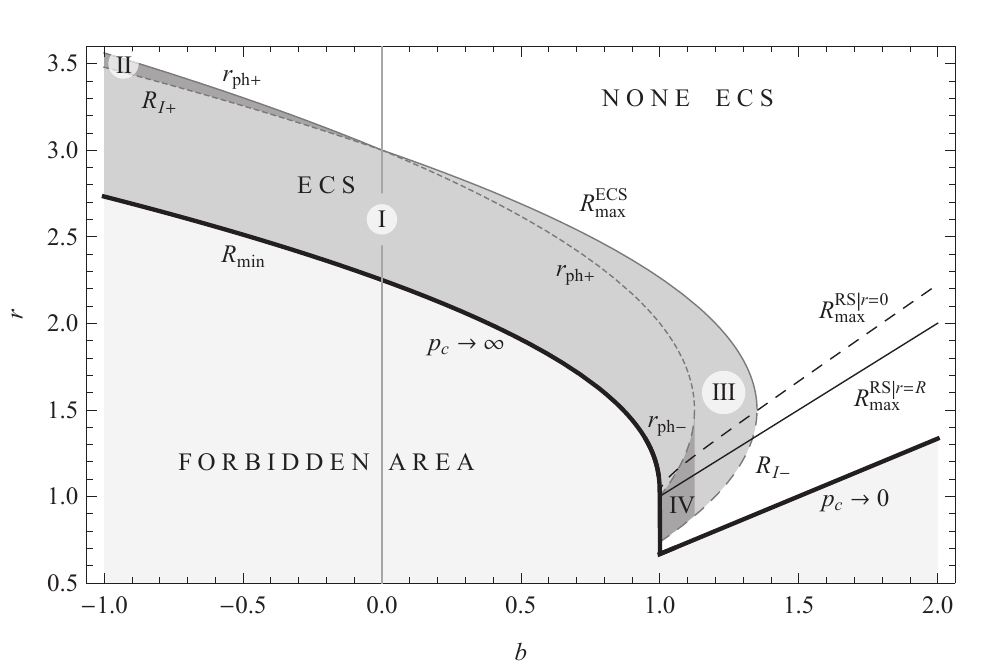}
      \caption{Classification of the compact stars. For the extremely compact stars (ECS), the parameter space $(b-R)$ is separated into four Zones~I--IV; their properties are given in the text.}\label{figure1}
    \end{figure}

    The reality condition of the central pressure reads $b + (R - 2) R\geq 0$ being equivalent to the condition $\Delta(R)\in \mathbb{R}$.
    The pressure in the center of the star must be finite and positive implying the limits $R_\mathrm{min}(b)$ on the existence of uniform density stars. The central pressure $p(r = 0, R, b)$ diverges when the surface radius $R$ satisfies the condition
    \begin{equation}
      3R \left(\sqrt{b + (R - 2) R} + R - 3\right) + 6b = 0
    \end{equation}
    that leads to a cubic equation relative to $R$ giving one solution that is relevant for $b\leq 1$ and reads
    \begin{eqnarray}
       R & = & \frac{1}{4} \left(\frac{(b - 9) (b - 1)}{\sqrt[3]{8 \sqrt{(b - 1)^2 b^3}+b \left[b (b + 17) - 45\right] + 27}}\right. + \qquad \qquad \qquad  \nonumber \\
         &   &\qquad \qquad \qquad \left.+ \sqrt[3]{8 \sqrt{(b - 1)^2 b^3} + b \left[b (b + 17) - 45\right] + 27} + b + 3\right).
    \end{eqnarray}
    The solution is depicted on Figure~\ref{figure1} as the part of $R_\mathrm{min}(b)$ denoted by $p \rightarrow \infty$. For $b = 0$, we arrive at the standard condition $R > \frac{9}{4}$ (see~\cite{Stu-Tor-Hle-Urb:2009:}). For $b > 1$ the relevant limit is given by the condition of positiveness of the central pressure. It reads
    \begin{equation}
      R_\mathrm{min}(b) = \frac{2}{3}b
    \end{equation}
    and is denoted by $p \rightarrow 0$. For smaller $R$ the central pressure is not positive.

    Note that for $3/4 < b < 1$ there is a region of surface radii given by the condition $2b/3 < R < 1 - \sqrt{1 - b}$ where $p(r = 0)$ is positive. However, the star surface is located under the inner horizon of the external spacetime that belongs to the black-hole type RN spacetimes. Such a configuration has to be hidden under the inner horizon of the external RN spacetime and is irrelevant for our considerations.

  \subsection{Extremely compact stars}\label{SECecs}
    The extremely compact stars (neutron, quark, or hybrid) are defined as compact stars allowing for existence of trapped null geodesics in their interior \cite{Stu-Tor-Hle-Urb:2009:,Stu-Hla-Urb:2011:}. In the case of internal uniform density Schwarzschild spacetimes ($b = 0$), such objects appear just when the surface of the compact star is located under the photon circular geodesic of the external vacuum Schwarzschild spacetime located at $r_\mathrm{ph} = 3r_\mathrm{g}$ \cite{Stu-Tor-Hle-Urb:2009:}. In the braneworld uniform density compact stars this simple rule does not hold and the situation is more complex due to different character of the external spacetime and its relation to the internal spacetime caused by the bulk space tidal effect on the matching of the spacetimes reflected by a non-standard way of the relations of the effective potential of the null geodesics in the internal and external spacetimes of extreme compact stars \cite{Stu-Hla-Urb:2011:}.

    The motion along null-geodesics is independent of energy (frequency) and can conveniently be described in terms of the impact parameter
    \begin{equation}
      \ell = \frac{L}{E},
    \end{equation}
    where $E$ and $L$ are the constants of motion related to the Killing vector fields of the internal and external spacetimes.
    The relevant equation governing the radial motion along null geodesics takes the form~\cite{Stu-Tor-Hle-Urb:2009:}
    \begin{equation}
      (p^{r})^{2} = A^{-2}(r)B^{-2}(r)E^{2} \left(1 - A^{2}(r)\frac{\ell^{2}}{r^{2}}\right).
    \end{equation}
    The energy $E$ is irrelevant and can be used for rescalling of the affine parameter $\lambda$. The radial motion is restricted by an effective potential related to the impact parameter $\ell$, defined by the relations
    \begin{equation}
      \ell^{2} \leq V{}_{\mathrm{eff}} =\left\{
      \begin{array}{lll}
        V{}_{\mathrm{eff}}^{\mathrm{int}} = \displaystyle\frac{r^2}{(A^{-}(r))^2}  & \quad\mbox{for} & r\leq R,\\
        V{}_{\mathrm{eff}}^{\mathrm{ext}} = \displaystyle\frac{r^2}{(A^{+}(r))^2}  & \quad\mbox{for} & r > R.
      \end{array}\right.
    \end{equation}
    $V{}_{\mathrm{eff}}^{\mathrm{int}}$ is the effective potential of the null-geodetical motion in the internal  braneworld spacetime and $V{}_{\mathrm{eff}}^{\mathrm{ext}}$ is the effective potential of the null-geodetical motion in the external, vacuum RN spacetime. Due to the bulk space tidal effects on the matching of the internal and external spacetimes (see \cite{Ger-Maar:2001:} for details) we obtain a non-standard variety of relations of the effective potentials in the ECS interior and exterior.

    Using the dimensionless radial coordinate expressed in terms of the gravitational radius ($r/G\mathcal{M} \rightarrow r$), and dimensionless tidal charge $b$, we obtain the relations

    \begin{equation}
      V{}_{\mathrm{eff}}^{\mathrm{int}} = \frac{r^2 R^2 \left[b Y - 2R (2b - 3R) Z\right]^2}{9 \left[b + (R - 2) R\right] \left[(R - b) + R (2b -3R) Z\right]^2},
    \end{equation}
    \begin{equation}
      V{}_{\mathrm{eff}}^{\mathrm{ext}} = \frac{r^4}{b + (r - 2) r},
    \end{equation}
    where
    \begin{eqnarray}
      Y &\equiv & R^2 \Delta(r) = \sqrt{r^2 (b - 2R) + R^4},\\
      Z &\equiv & R \Delta(R) = \sqrt{b + (R - 2) R}.
    \end{eqnarray}

    Circular null geodesics, located at $r_\mathrm{c(i)}$ and $r_\mathrm{c(e)}$, respectively, are given by the local extrema of the effective potential ($\partial V_\mathrm{eff}/\partial r = 0$). In the internal spacetime we have to solve a nontrivial equation~\cite{Stu-Hla-Urb:2011:}:
    \begin{equation}
      \begin{array}{l}
        2 r R^2 \left[b (Y - 4 R Z) + 6 R^2 Z\right] \left\{-b^3 \left[r^2 (Y - 4 R Z) + 8 R^2 Y\right] + b^2 R \left[r^2 (3 Y - 14 R Z) +  R^2 \times \right.\right. \\
        \left.\times \left(6 R^2 Z - 9 R Y + 40 Y\right)\right] + b R^2 \left[R^2 \left(-13 R^2 Z + 25 R Y - 66 Y\right) - 2 r^2 (Y - 6 R Z) \right] + 6 R^5 \times \\
        \left.\left. \times \left(R^2 Z - 3 R Y + 6 Y\right)\right\}\right]
        \left[9 \left\{b + (R - 2) R\right\} Y \left\{R (Y - 3 R Z) - b \left(Y - 2 R Z\right)\right\}^3\right]^{-1} = 0.
      \end{array}
    \end{equation}

    This equation can be solved by using numerical methods and determines the loci $r_\mathrm{c(i)}(R, b)$ of the internal circular null geodesics that are stable and correspond to a local maximum of the internal effective potential (see~\cite{Stu-Hla-Urb:2011:} for details). The existence of the internal stable circular null geodesics is allowed only for the surface radius $R$ limited by values explicitly given by
    \begin{equation}
       R_\mathrm{I\pm} = \frac{3}{2} \pm \sqrt{\frac{9}{4} - \frac{5}{3}b}
    \end{equation}
    that has to satisfy naturally also the condition $R > R_\mathrm{min}(b)$.

    In the external spacetime, the photon circular geodesics and the related local extrema of the effective potential of the Reissner-Nordstr\"{o}m (RN) type, with the tidal charge substituting the electric charge squared appearing in the standard RN spacetimes, are given in the dimensionless units by the condition
    \begin{equation}
      r^2 - 3r + 2b = 0.
    \end{equation}
    For negative tidal charges ($b < 0$) external spacetimes of the black-hole type are allowed only~\cite{Kot-Stu-Tor:2008:CLASQG:}.
    The radii of the photon circular orbits are given by~\cite{Bal-Bic-Stu:1989:,Stu-Hle:2002:}
    \begin{equation}
      r_{\mathrm{ph}\pm} = \frac{3}{2}\left(1 \pm \sqrt{1 - \frac{8}{9}b}\right). \label{EQph}
    \end{equation}
    However, only the outer photon circular geodesic at $r_\mathrm{ph+}$ is physically relevant and we see immediately that for $b < 0$ there is $r_{\mathrm{ph}}>3$.

    \begin{figure}[t]%
      \begin{minipage}[b]{.499\hsize}
        \centering\includegraphics[width=\hsize,keepaspectratio=true]{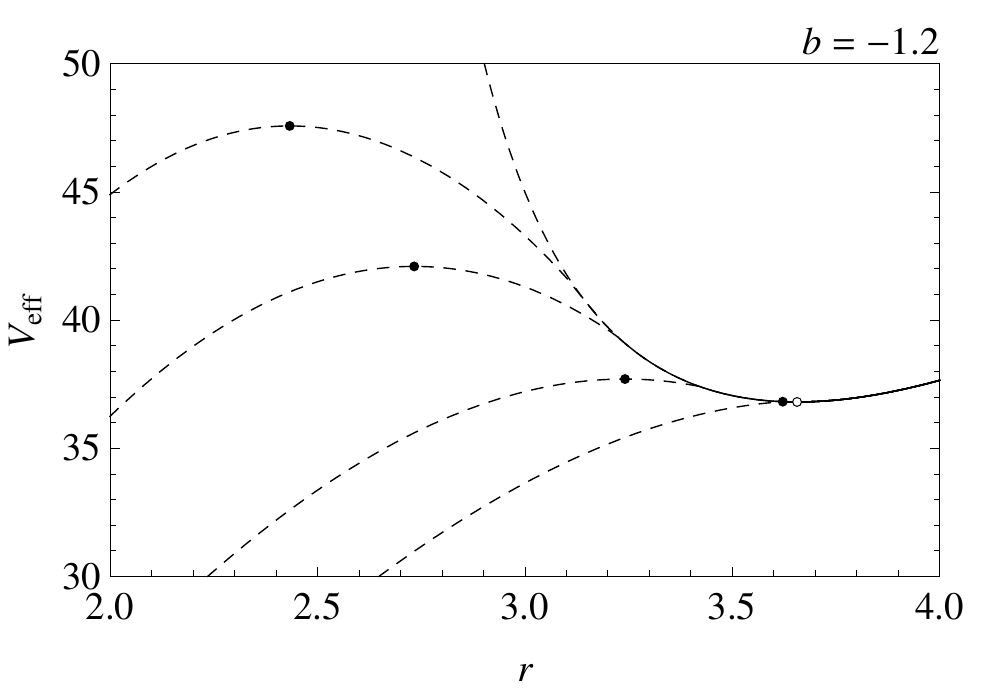}
        \centering\includegraphics[width=\hsize,keepaspectratio=true]{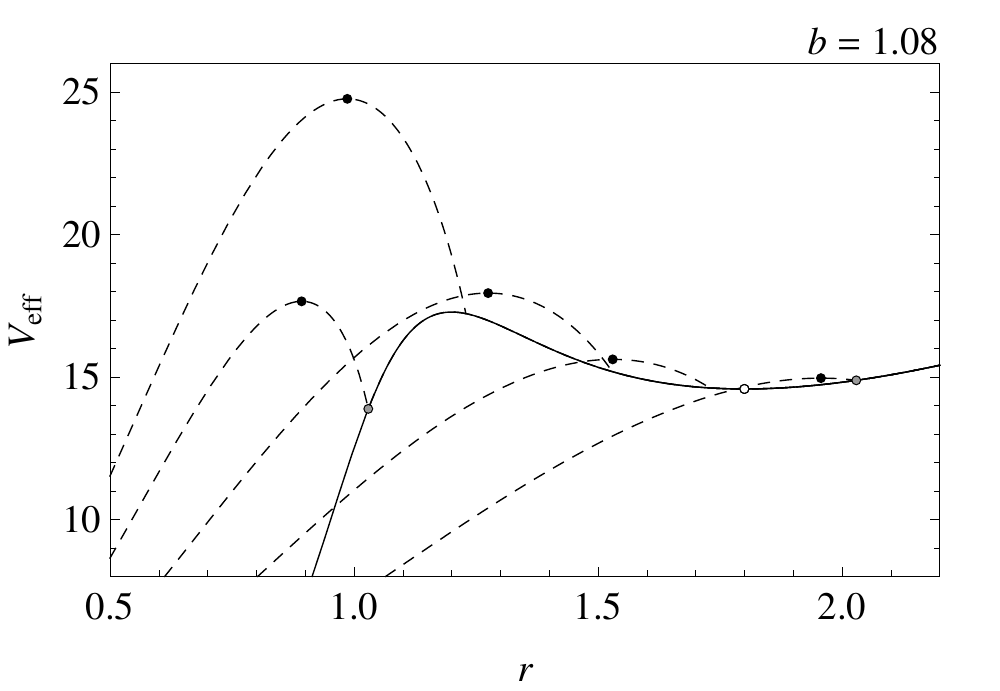}
      \end{minipage}\hfill%
      \begin{minipage}[b]{.499\hsize}
        \centering\includegraphics[width=\hsize,keepaspectratio=true]{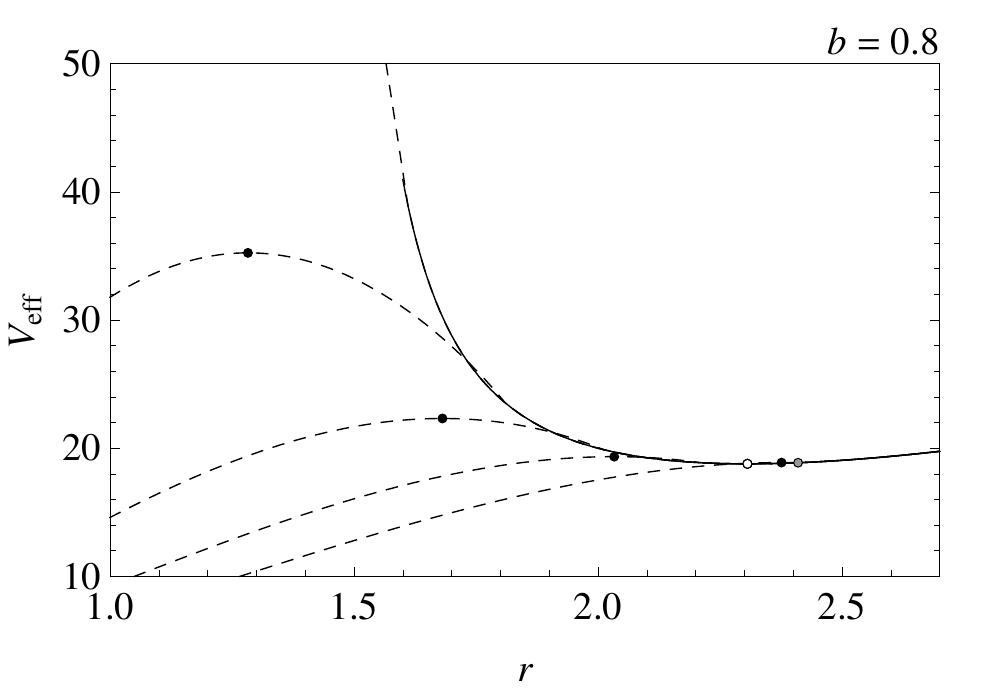}
        \centering\includegraphics[width=\hsize,keepaspectratio=true]{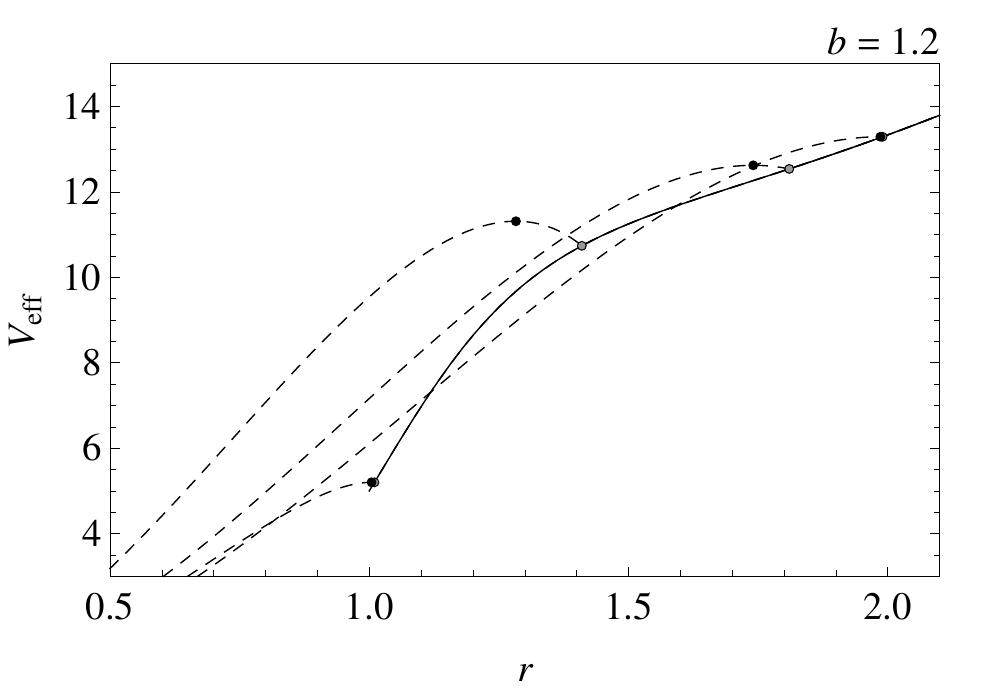}
      \end{minipage}
      \caption{The effective potential $V_{\mathrm{eff}}$ of both internal (dashed lines) and external spacetimes (full lines) given for the black-hole type ($b < 1$) and naked-singularity ($b > 1$) type of the external spacetime. All characteristic cases of its behaviour are presented for both ECS and non-extreme compact stars.}\label{figure2}
    \end{figure}

    For braneworld neutron stars with a positive tidal charge ($b > 0$) both black-hole and naked-singularity RN spacetimes are relevant \cite{Kot-Stu-Tor:2008:CLASQG:}. In the black-hole type spacetimes ($b \leq 1$), the photon circular orbits are given by the relation~\eqref{EQph}, but only the outer solution corresponding to an unstable orbit can be astrophysically relevant since the surface of the compact star has to be located above radius of the RN outer horizon. In the naked-singularity spacetimes ($b > 1$) two photon circular orbits can be astrophysically relevant when $b < 9/8$, the inner one being stable, the outer one --- unstable. In the spacetimes with $b > 9/8$, no photon circular orbits can exist. Therefore, in the braneworld spacetimes, the existence of ECS is governed by the interplay of the behaviour of the internal and external effective potential of the motion and will be determined and classified in the following way.

    When we take into account the (non-)existence of the local extrema of the internal/external part of the effective potential $V_\mathrm{eff}$, we obtain the whole region of existence of ECS.
    The function $R{}_\mathrm{max}^\mathrm{ECS}(b)$ is determined by the relation
    \begin{equation}
      R{}_\mathrm{max}^{\mathrm{ECS}} =\left\{
        \begin{array}{lll}
          r_\mathrm{ph} \equiv \frac{1}{2} \left(3 + \sqrt{9 - 8 b}\right) & \quad\mbox{for} & b \leq 0,\\
          R_\mathrm{I+} \equiv \frac{3}{2} + \sqrt{\frac{9}{4} - \frac{5}{3}b}  & \quad\mbox{for} & 0 < b < 27/20,
        \end{array}
      \right.
    \end{equation}
    while for $R{}_\mathrm{min}^\mathrm{ECS}(b)$ we have
    \begin{equation}
      R{}_\mathrm{min}^{\mathrm{ECS}} = \left\{
        \begin{array}{lll}
          R_\mathrm{min} & \quad\mbox{for} & b \leq 1,\\
          R_\mathrm{I-} \equiv \frac{3}{2} - \sqrt{\frac{9}{4} - \frac{5}{3}b} & \quad\mbox{for}& 1 < b < 27/20.
        \end{array}
      \right.
    \end{equation}
    The region of ECS in the parameter space is represented in Figure~\ref{figure1}.

    Typical behaviour of the effective potential of the null-geodetical motion $V{}_{\mathrm{eff}}$ is demonstrated in Figure~\ref{figure2} for appropriately chosen values of the tidal charge $b$. Here we express the radii in units of the gravitational radius $r_\mathrm{g} = G \mathcal{M}$. The selection of the tidal charge values used in Figure~\ref{figure2} demonstrates the full classification of the behaviour of the effective potential in both internal and external spacetimes. When the effective potential in the internal or external spacetime (or in both of them) has a local extreme corresponding to a photon circular geodesic, trapped null geodesics can appear if the surface radius $R$ is properly chosen giving an ECS. On the other hand, we find an ordinary compact star if the surface is chosen in such a way that no local extrema of the effective potentials exist (see Figure~\ref{figure2}). Separation of the zones of compact stars of different character, both extremely compact and ordinary compact, in the parameter space $(b-R)$, is determined by the functions $R_\mathrm{min}(b)$, $R_\mathrm{max}(b)$ and $r_\mathrm{ph}(b)$ where the last function governs radius of the photon circular geodesics in the external spacetime (both of them for $1 < b < 9/8$). The subdivision of the Zone IV in the tidal charge range $1 < b < 9/8$ is given by the relation of the magnitude of the effective potential at the surface of the ECS and the magnitude of the external effective potential at its local minimum, and is determined numerically.

    The parameter space $R-b$ is divided into regions corresponding to the existence of four different zones of ECS. The classification is based on the existence of local maxima/minima of $V{}^{\mathrm{int/ext}}_{\mathrm{eff}}$ in the following way (see Figures~\ref{figure1} and \ref{figure2}):
    \begin{itemize}
      \item[$\bullet$]{\emph{Zone I} there exist maximum of $V{}^{\mathrm{int}}_{\mathrm{eff}}$ and minimum of $V{}^{\mathrm{ext}}_{\mathrm{eff}}$, both located at $r \neq R$, $\mathrm{min} V{}^{\mathrm{ext}}_{\mathrm{eff}} < V{}^{\mathrm{ext}}_{\mathrm{eff}}(r = R)$, and there is no local maximum of $V{}^{\mathrm{ext}}_{\mathrm{eff}}$ at $r > R$;}
      \item[$\bullet$]{\emph{Zone II} there exist maximum of $V{}^{\mathrm{int}}_{\mathrm{eff}}$ at $r = R$  and minimum of $V{}^{\mathrm{ext}}_{\mathrm{eff}}$ at $r \neq R$, $\mathrm{min} V{}^{\mathrm{ext}}_{\mathrm{eff}} < \mathrm{max} V{}^{\mathrm{int}}_{\mathrm{eff}}$, and there is no local maximum of $V{}^{\mathrm{ext}}_{\mathrm{eff}}$ at $r > R$;}
      \item[$\bullet$]{\emph{Zone III} there exist maximum of $V{}^{\mathrm{int}}_{\mathrm{eff}}$ at $r < R$  and minimum of $V{}^{\mathrm{ext}}_{\mathrm{eff}}$ at $r = R$; there is no local maximum of $V{}^{\mathrm{ext}}_{\mathrm{eff}}$ at $r > R$;}
      \item[$\bullet$]{\emph{Zone IV} there exist both the local maximum and local minimum of $V{}^{\mathrm{ext}}_{\mathrm{eff}}$ at $r > R$. The last zone can be divided into two parts according to the criterion~\cite{Stu-Hla-Urb:2011:}
          \begin{itemize}
            \item[a)]{minimum of $V{}^{\mathrm{ext}}_{\mathrm{eff}} < V{}^{\mathrm{int}}_{\mathrm{eff}}\left(r = R\right)$,}
            \item[b)]{ minimum of $V{}^{\mathrm{ext}}_{\mathrm{eff}} > V{}^{\mathrm{int}}_{\mathrm{eff}}\left(r = R\right)$.}
          \end{itemize}}
    \end{itemize}

  \section{Redshift of radiation from the braneworld RN compact stars}\label{SECredsh}
    We can find interesting restrictions on the parameters of the braneworld compact stars considering redshift of radiation; electromagnetic radiation emitted from the surface of the compact stars, and neutrino radiation emitted from the interior of cooling compact stars, assuming compact stars cooled down enough to allow the approximation of the geodesic motion of neutrinos. Of course, we have to expect development of sophisticated observational techniques enabling measurement of the energy distribution in the case of neutrinos radiated by newly born neutron stars that cooled down sufficiently to allow for the neutrino free motion in the whole star interior.

    The geodetical motion of photons in the external vacuum spacetime is fully justified, contrary to the neutrino motion in the neutron star interior. The approximation of free motion of neutrinos in the internal spacetime is correct when the mean free path of neutrinos $\lambda > R$. Neutrinos have inelastic scatter on electrons (muons) and elastic scatter on neutrons. The scatter cross section on electrons (neutrons) $\sigma_\mathrm e$ ($\sigma_\mathrm n$) gives the mean free path in the form $\lambda = (\sigma_i n_i)^{-1}$ where $n_i$ $(i\in \{\mathrm{e, n}\})$ denotes the number density of electrons (neutrons). It was shown \cite{Sha-Teu:1983:BHWDNS:} that
    \begin{equation}
      \lambda_\mathrm{e}\sim 9\times 10^7\left(\frac{\rho_\mathrm{nucl}}{\rho} \right)^{4/3}\left(\frac{100\mathrm{keV}}{E_\nu}\right)^{3}~\mathrm{km},
    \end{equation}
    while
    \begin{equation}
      \lambda_\mathrm{n}\sim 300\frac{\rho_\mathrm{nucl}}{\rho} \left(\frac{100\mathrm{keV}}{E_\nu}\right)^{2}~\mathrm{km}.
    \end{equation}
    There is $\lambda_\mathrm e \gtrsim  10$~km for $E_\nu \lesssim 20$~MeV and $\lambda_\mathrm n \gtrsim 10$~km for $E_\nu \lesssim 500$~keV. Therefore, in a few hours old neutron star, see \cite{Lat-Pra:2007:PhysRep:,Sha-Teu:1983:BHWDNS:,Gle:2000:CompactStars:,Web:1999:Pul:}, at temperatures $T \lesssim 10^9$~K ($E_\nu \sim 100$~keV), the neutrino motion could be considered geodetical through whole the internal spacetime. Of course, our results can be applied at any time of the neutrino radiation as they give the upper limit on the observed redshift of neutrinos.

    It is usual to make estimates on the redshift of radiation of the surface and interior of the compact stars using the simplest possibility of the purely radial motion of photons and neutrinos. The redshift factor related to the neutron star interior or surface is given (under the assumption of a static source and radially emitted photons and neutrinos) by the standard formula $\mathrm{d}\tau^2 = -g_{tt} \mathrm{d}t^2$ relating proper time of the source ($\tau$) and the observer ($t$) that implies
    \begin{equation}
      \frac{\nu_\mathrm{e}}{\nu_\mathrm{o}} = \frac{1}{\sqrt{-g_{tt}}} = 1 + z.
    \end{equation}

\begin{figure}[t]
      \centering\includegraphics[width=0.6\hsize]{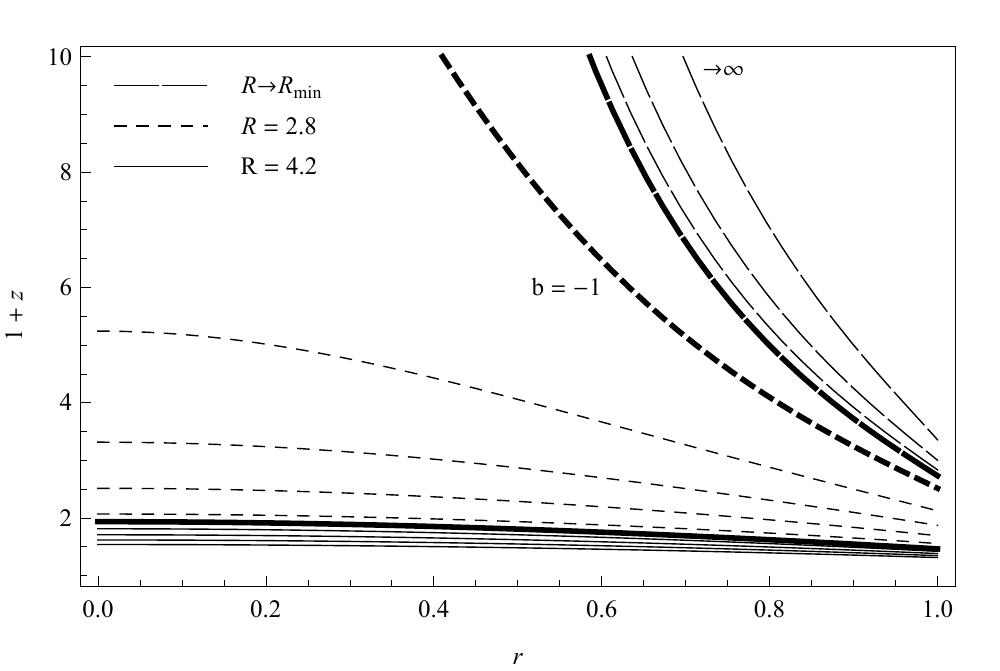}
      \caption{The dependence of $1 + z$ on the brane parameter $b \in \{-1, -0.5, 0, 0.5, 1)$ and $r$ is demonstrated for several values of $R$. Curves for $b = -1$ are plotted thick. For $R \rightarrow R_\mathrm{min}$ the central redshifts diverge; this effect is amplified with $b$ getting closer to $1$, even so much, that there is no curve on the plot for case $b \rightarrow 1$, $R \rightarrow R_\mathrm{min}(b \rightarrow 1)$.}\label{figure3}
    \end{figure}

    In Figure~\ref{figure3} we give an illustration of the redshift profile in dependence on the tidal charge $b$ and the surface radius $R$ in the interval corresponding to the lower limit of the radius of extremely compact stars $R_\mathrm{min}$ when the redshift diverges at the centre and some relatively large and astrophysically acceptable radius $R = 4.2$ representing relatively flat profile related to realistic compact stars.

    We first study in detail the redshift of photons radiated from the surface of the braneworld compact stars.

  \subsection{Surface redshift}
    The redshift factor related to the neutron star surface implies
    \begin{equation}
       \frac{\nu_e}{\nu_o} = (1 + z)(r = R)  = \frac{1}{\sqrt{-g_{tt}(r = R)}} .
    \end{equation}

    \begin{figure}[t]%
      \begin{center}
        \includegraphics[width=.499\hsize]{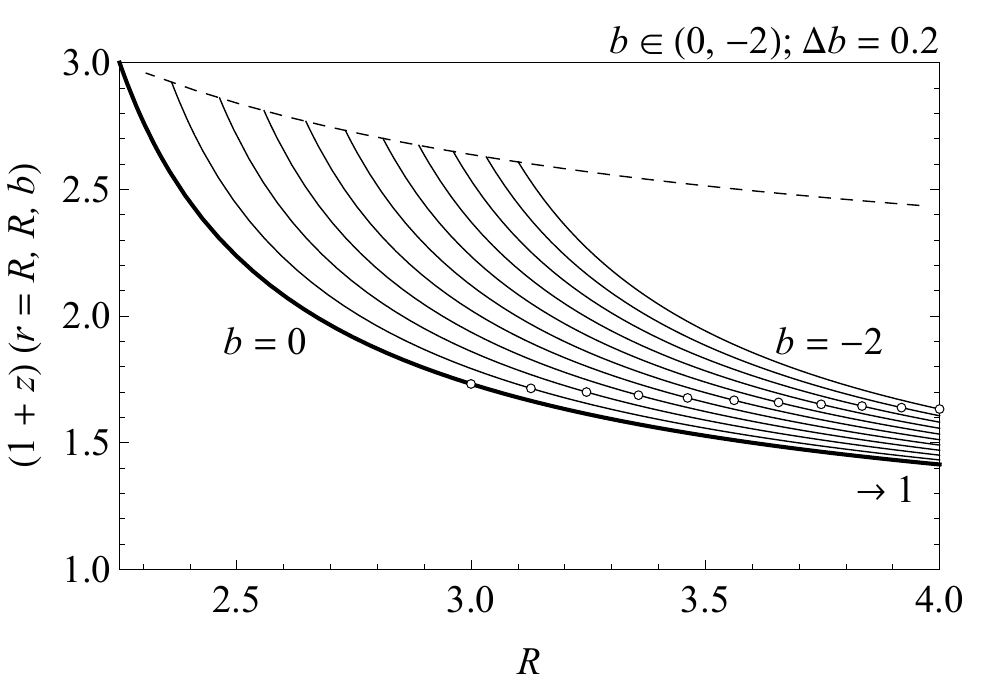}\hfill\includegraphics[width=.499\hsize]{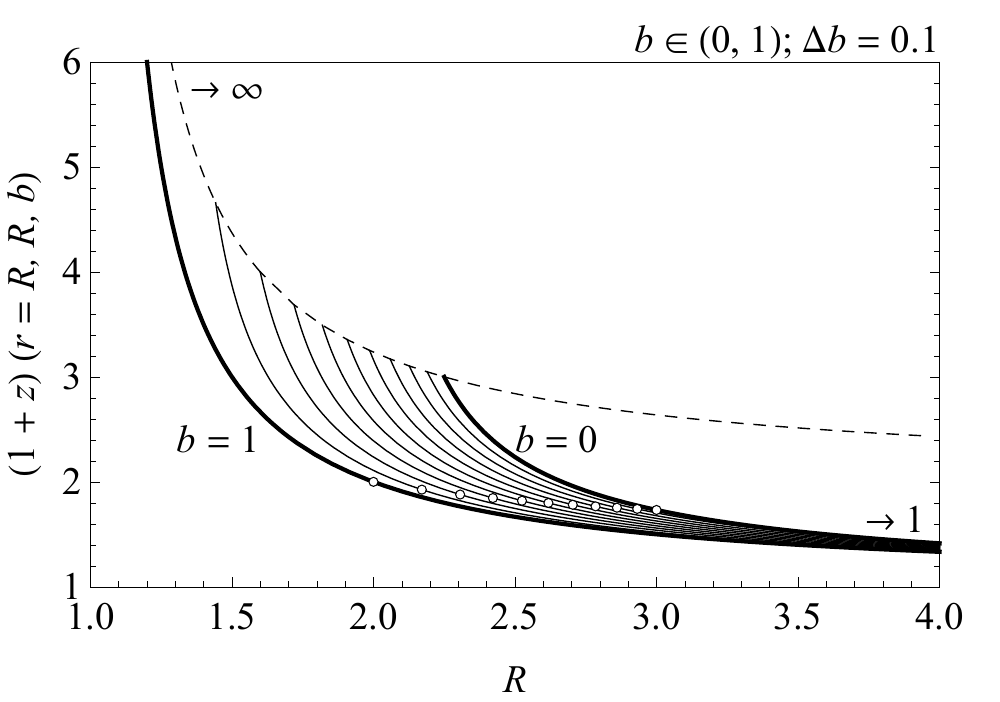}
      \end{center}
      \centering\includegraphics[width=.499\hsize]{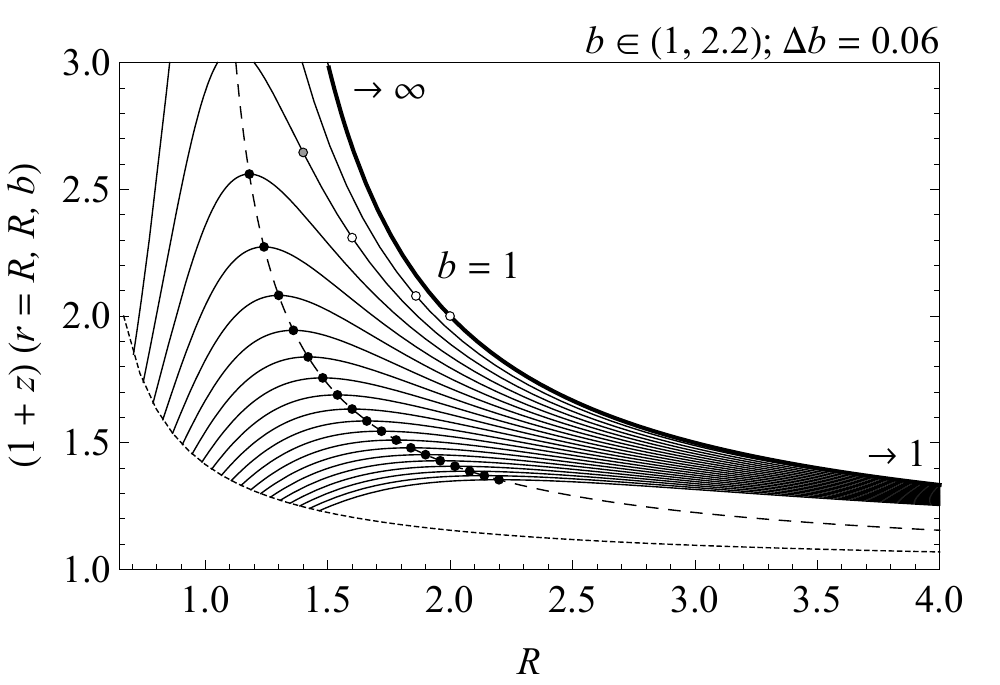}
      \caption{Surface redshifts of the compact stars given for typical values of the tidal charge --- for negative ($b < 0$) and positive ($b > 0$) tidal charge spacetimes of the black-hole type and the $b > 1$ case of naked-singularity type external spacetime.}\label{figure4}
    \end{figure}

    The external geometry at the surface radius
    \begin{equation}
      -g_{tt}(r = R) = \left[1 - 2\frac{r_\mathrm{g}}{R} + b\left(\frac{r_\mathrm{g}}{R}\right)^2\right].
    \end{equation}
    Using the dimensionless units $r_\mathrm{g} = 1$ here and in the following, the surface redshift factor reads
    \begin{equation}
      1 + z = \left[1 - \frac{2}{R} + \frac{b}{R^2}\right]^{-1/2},
    \end{equation}
    where the condition $R > R_\mathrm{min}(b)$ must be satisfied. Behaviour of the surface redshift is illustrated in Figure~\ref{figure3}, where the loci $R = r_\mathrm{ph}$ are depicted by circles. The surface redshift is given in Figure~\ref{figure4} for negative and positive tidal charges with surface radius going down to $R = R_\mathrm{min}$. For the surface located at the unstable photon circular orbit ($R = r_\mathrm{ph+}$) the redshift slightly decreases (increases) with descending $b < 0$ (increasing $b > 0$). For $R \rightarrow R_\mathrm{min}(b)$, the surface redshift slightly decreases with decreasing $b < 0$, but it substantially increases with increasing $b > 0$ and diverges for $b \rightarrow 1$.

    For positive tidal charges  $b > 1$ corresponding to naked-singularity spacetimes the redshift radial profile reaches a maximum at $R = b$ where
    \begin{equation}
      (1 + z)_\mathrm{max}(r = R = b) = \sqrt{\frac{b}{b - 1}};
    \end{equation}
    its position is depicted in Figure~\ref{figure4} by black points. For $b\leq 9/8$ we depict also the radius corresponding to photon circular orbits.

    It is important to determine the redshift of the braneworld neutron stars with surface located at the radii corresponding to the photon circular geodesics of the external spacetime --- in the case of the standard internal Schwarzschild spacetimes with $b = 0$ the case of limiting (maximally extended) extremely compact objects coincides with their surface being located at $R = r_{\mathrm{ph}} = 3$. Then we find the well known value of the surface redshift
    \begin{equation}
      (1 + z)(r = R = r_{\mathrm{ph+}}, b=0) = \left(1 - \frac{2}{3}\right)^{-1/2} = \sqrt{3}.
    \end{equation}

    \begin{figure}[t]%
      \centering\includegraphics[width=0.7\hsize]{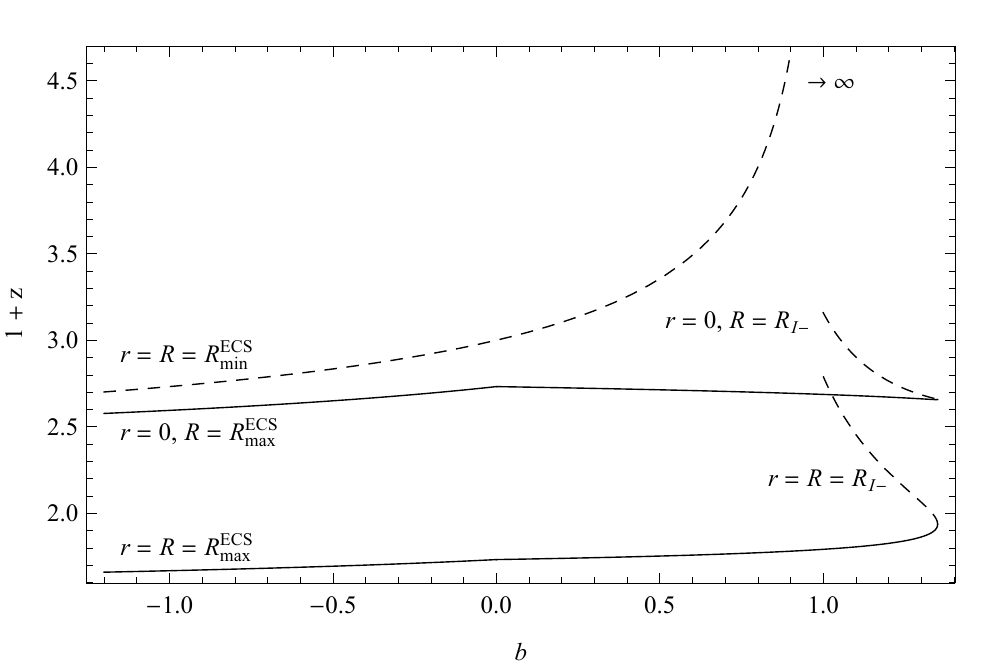}
      \caption{The dependence of surface redshift and central redshift on the brane parameter $b\in(-1.2,27/20)$ for ECS.}\label{figure5}
    \end{figure}

    On the other hand, in the braneworld case with $b \neq 0$ we have to distinguish the black-hole type external spacetimes with one photon circular orbit, and the naked-singularity type spacetimes when two photon circular orbits exist --- the internal one being stable, the external one being unstable. It is important to distinguish these cases because of the qualitatively different character of optical phenomena at the external field with strong influence on the accretion discs \cite{Stu-Schee:2010:CLAQG:,Stu-Hle:2000:} that could be astrophysically very important and could provide an independent evidence on the tidal charge presence. The surface redshift at $R = r_\mathrm{ph}$ in the black-hole type spacetimes is given by the relation
    \begin{equation}
      (1 + z)(r = R = r_\mathrm{ph}, b) = 2 \sqrt{\frac{2b}{4b - 3 + \sqrt{9 - 8b}}}.
    \end{equation}
    This dependence is illustrated in Figure~\ref{figure4}.
    In the naked-singularity-type spacetimes admitting existence of the photon circular orbits, i.e., for $1 < b < 9/8$, the surface redshift is determined by
    \begin{equation}
      (1 + z)(r = R = r_\mathrm{ph\pm}, b) = 2\, \sqrt{\frac{2b}{4 b\pm\sqrt{9 - 8 b} - 3}},
    \end{equation}
    and for both circular orbits is given in Figure~\ref{figure4}. For $b > 9/8$ no photon circular orbits exist.

    \begin{figure}[t]%
      \centering\includegraphics[width=0.7\hsize]{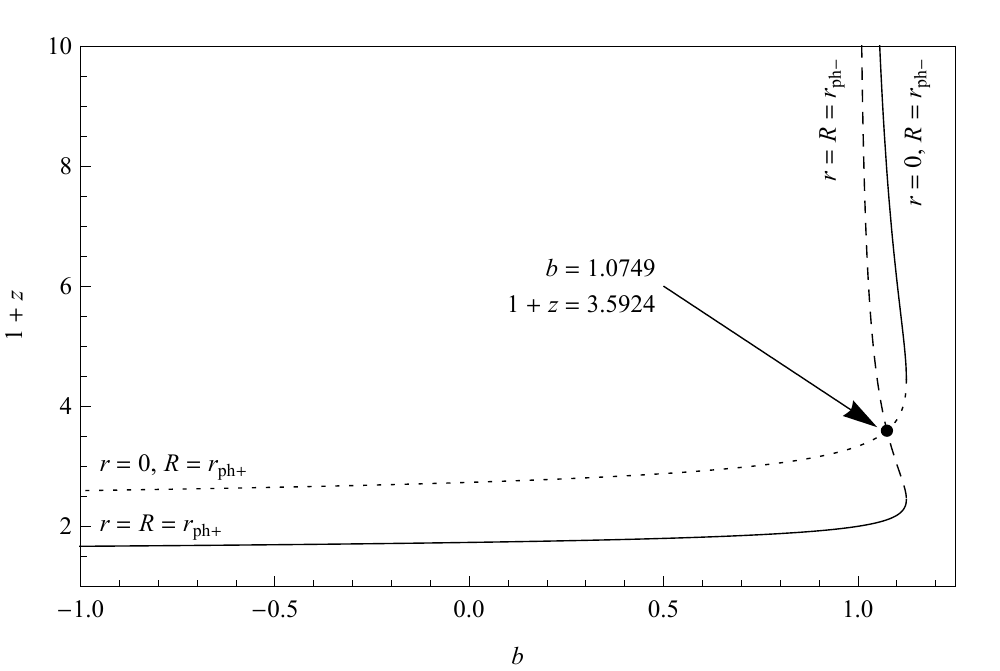}
       \caption{The figure illustrates the span of $1 + z$ for the interior of the star with $R = r_\mathrm{ph}$. }\label{figure6}
    \end{figure}

    For the limiting, minimally and maximally extended extremely compact braneworld spacetimes the value of surface redshift is varying in the range starting from $(1+z){}_\mathrm{min}^{\mathrm{ECS}}$ and finishing at $(1+z){}_\mathrm{max}^{\mathrm{ECS}}$, determined by relations
    \begin{equation}
      (1 + z){}_\mathrm{min}^{\mathrm{ECS}} =\left\{
        \begin{array}{lll}
          (1+z)(r = R = r_\mathrm{ph+}, b) & \quad\mbox{for} & b \leq 0,\\
          (1+z)(r = R = R_\mathrm{I+}, b) & \quad\mbox{for} & 0 < b < 27/20,
        \end{array}
      \right.
    \end{equation}

    \begin{equation}
      (1+z){}_\mathrm{max}^{\mathrm{ECS}} =\left\{
        \begin{array}{lll}
          (1+z)(r = R = R_\mathrm{min}, b) = \Delta(R_\mathrm{min})^{-1} & \quad\mbox{for} & b \leq 1,\\
          (1+z)(r = R = R_\mathrm{I-}, b) & \quad\mbox{for} & 1 < b < 27/20,
        \end{array}
      \right.
    \end{equation}
    where
    \begin{equation}
      (1+z)(r = R = R_\mathrm{I\pm}, b) = 5\, \sqrt{\frac{2b}{20 b\pm \sqrt{81 - 60 b} - 9}}.
    \end{equation}
    We give the dependencies of the surface redshift limiting values on the braneworld tidal charge parameter $b$ in Figures~\ref{figure4} and \ref{figure5}. For special cases, when the compact star surface is located at the radii corresponding to photon circular orbits of the external spacetime, the dependence of surface redshift on tidal charge is illustrated in Figure~\ref{figure6}.

  \subsection{Redshift factor of neutrinos radiated at the interior of the star}
    In principle, distribution of the internal redshift factor enables to estimate the internal structure of the neutron star if the energy (and frequency shift) of outcoming neutrinos can be measured precisely enough.

    The redshift factor gives frequency shift of the neutrinos radiated by static sources inside the star. We assume radially emitted neutrinos --- this implies some restrictions on the observations but we shall not consider here the effect of bending of null geodesics since it is not too relevant for the static sources in the compact star interior. Of course, in the case of the source at the center of the star, only radially emitted neutrinos are, in fact, allowed, putting thus a natural restriction on the observed range of the redshift.

    The general redshift formula for static sources and purely radial emission reads
    \begin{equation}
      1 + z = \frac{\nu_e}{\nu_o} = \left[-g_{tt}(r)\right]^{-1/2} = \left[A^{-}(r)\right]^{-1} =  \frac{1 + p(r)/\varrho}{\Delta(R)}.
    \end{equation}
    We can see immediately that $(1+z) \rightarrow \infty$  if $p(r) \rightarrow \infty$. This occurs for all $R \rightarrow R_\mathrm{min}(b < 1)$. On the other hand, for fixed parameters $R$ and $b$, the behaviour of $1+z$ is simply given by the behaviour of the pressure function $p(r)$ that can be studied from its partial derivative
    \begin{equation}
      \frac{\partial p(r)/\varrho}{\partial r}=\frac{r R Z (2 b - 3 R) (b - 2 R)^2}{3 Y \left[R (Y - 3 R Z) - b (Y - 2 R Z)\right]^2}.
    \end{equation}
    Because $r > 0$ and $R > 2b/3$ for all considered braneworld compact stars (see Figure~\ref{figure1}), there is $\partial p(r)/\partial r < 0$, and we can conclude that $(1+z)(r,R,b)$ is (for fixed $R$ and $b$) monotonously decreasing function of increasing $r$; therefore, the tidal effects does not change this standard and intuitively expected behaviour of the pressure (and redshift) profile.

    Using the internal metric coefficients expressed in terms of the external metric parameters, we arrive at the redshift expressed in the form
    \begin{equation}
      (1 + z)(r, R, b) =  \frac{2\Delta(R)\left(2b-3R\right)-\Delta(r)b}{3\Delta(R)\left[\Delta(R) \left(2b - 3 R\right)+\Delta(r)\left(R-b\right)\right]}.
    \end{equation}

    For a fixed $b > 1$ the function $(1 + z)(r = 0, R, b)$ has a maximum occurring for a specific value of surface radius $R$. Its value can be found numerically by solving equation
    \begin{equation}
    \begin{array}{l}
      8 b^4 + b^3 R \left[9 R - 4 \left(\sqrt{b + (R - 2) R} + 12\right)\right] + b^2 R^2 \left(9 \sqrt{b + (R - 2) R} - 33 R + 106\right) + b R^3 \times\\
      \times \left[R \left(-\sqrt{b + (R - 2) R} + R + 41\right) - 2 \left(2 \sqrt{b + (R - 2) R} + 51\right)\right] - 18 (R - 2) R^4 = 0.\qquad
    \end{array}
    \end{equation}
    Solution is given in Figure~\ref{figure1} as the curve $R{}_\mathrm{max}^{\mathrm{RS}|r=0}$.

    \begin{figure}[t]%
      \begin{center}
        \includegraphics[width=.499\hsize]{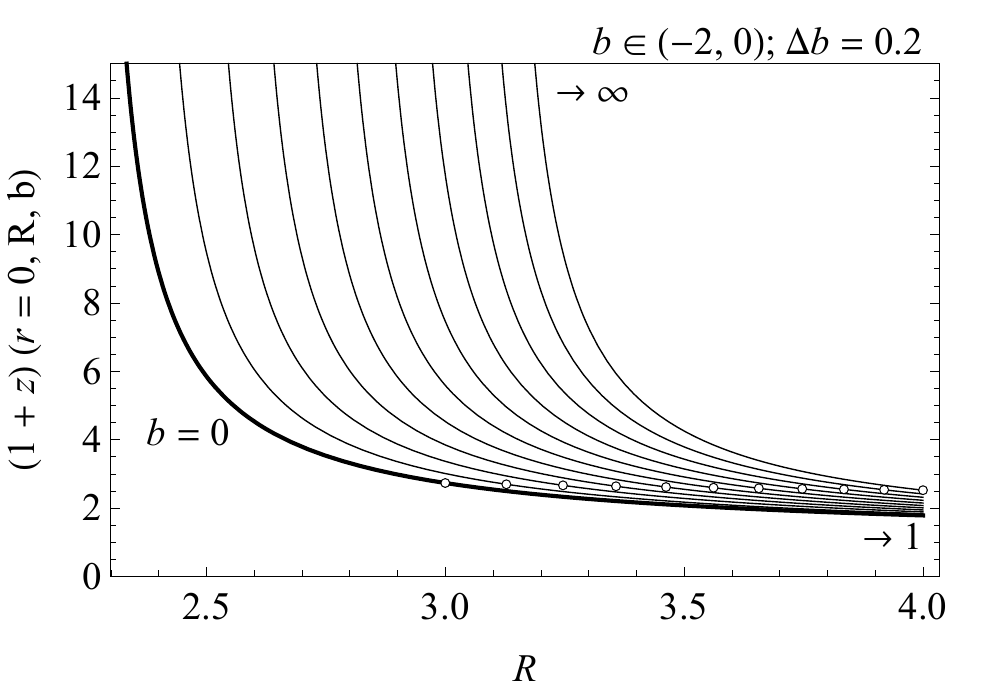}\hfill\includegraphics[width=.499\hsize]{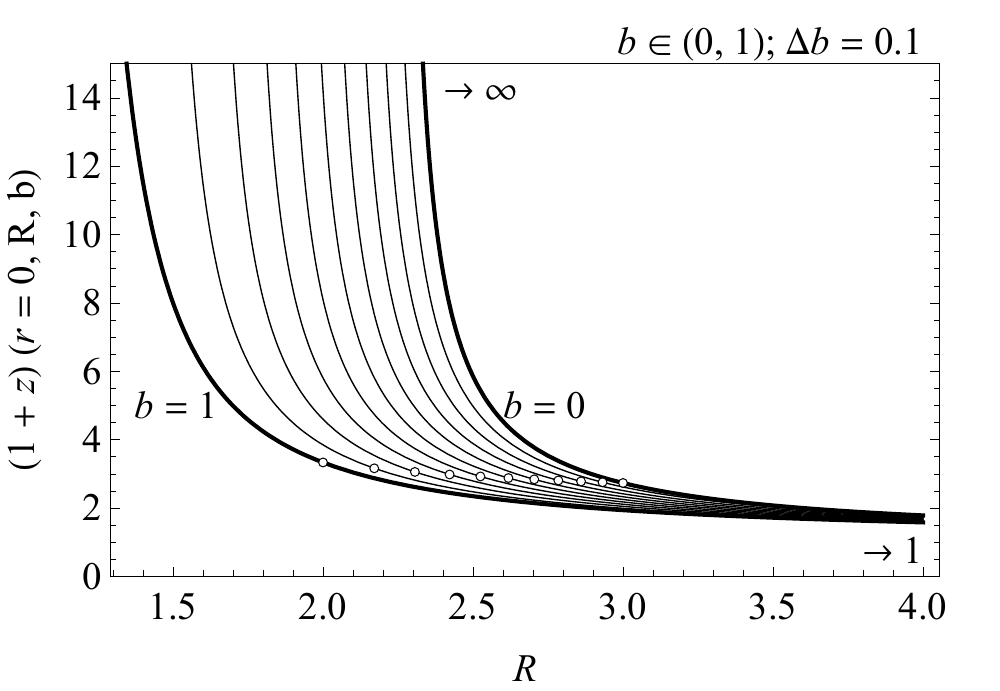}
      \end{center}
      \centering\includegraphics[width=.499\hsize]{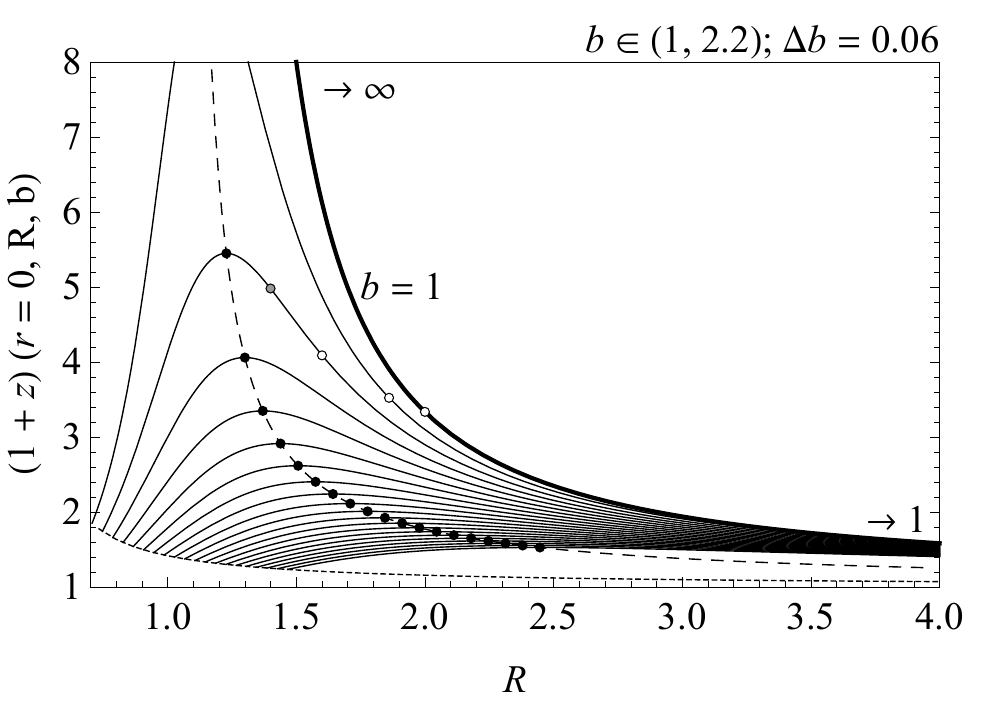}
      \caption{Plot of the redshifts $1 + z$ from the center of the star for several values of $b$.}\label{figure8}
    \end{figure}

    For the surface redshift at $r = R$, where $p(R) = 0$, we obtain the first limiting value of the redshift to be
    \begin{equation}
      (1 + z)(r = R, R, b) = \Delta(R)^{-1} = \left[1 - 2\frac{1}{R} + b\left(\frac{1}{R}\right)^2\right]^{-1/2}.
    \end{equation}

    For the fixed $R$ and $b$, the second limiting value of the redshift is given by $(1 + z)(r = 0, R, b)$ i.e., by its value at the center of the star. Since $\Delta(r = 0) = 1$, we find
    \begin{equation}
      (1 + z)(r = 0, R, b) = \frac{2\Delta(R)(3R - 2b) + b}{3\Delta(R)\left[\Delta(R)(3R - 2b) + (b - R)\right]}.
    \end{equation}
    In the standard internal Schwarzschild spacetime ($b = 0$), the central redshift reads
    \begin{equation}
      (1 + z)(r = 0, R, b = 0) = \frac{2}{3 \sqrt{1 - 2/R} - 1}
    \end{equation}
    and for the maximally extended ECS there is
    \begin{equation}
      (1 + z)(r = 0, R = 3, b = 0) = 1 + \sqrt{3}.
    \end{equation}

\begin{figure}[t]%
      \begin{minipage}[b]{.445\hsize}
        \centering\includegraphics[width=\hsize,keepaspectratio=true]{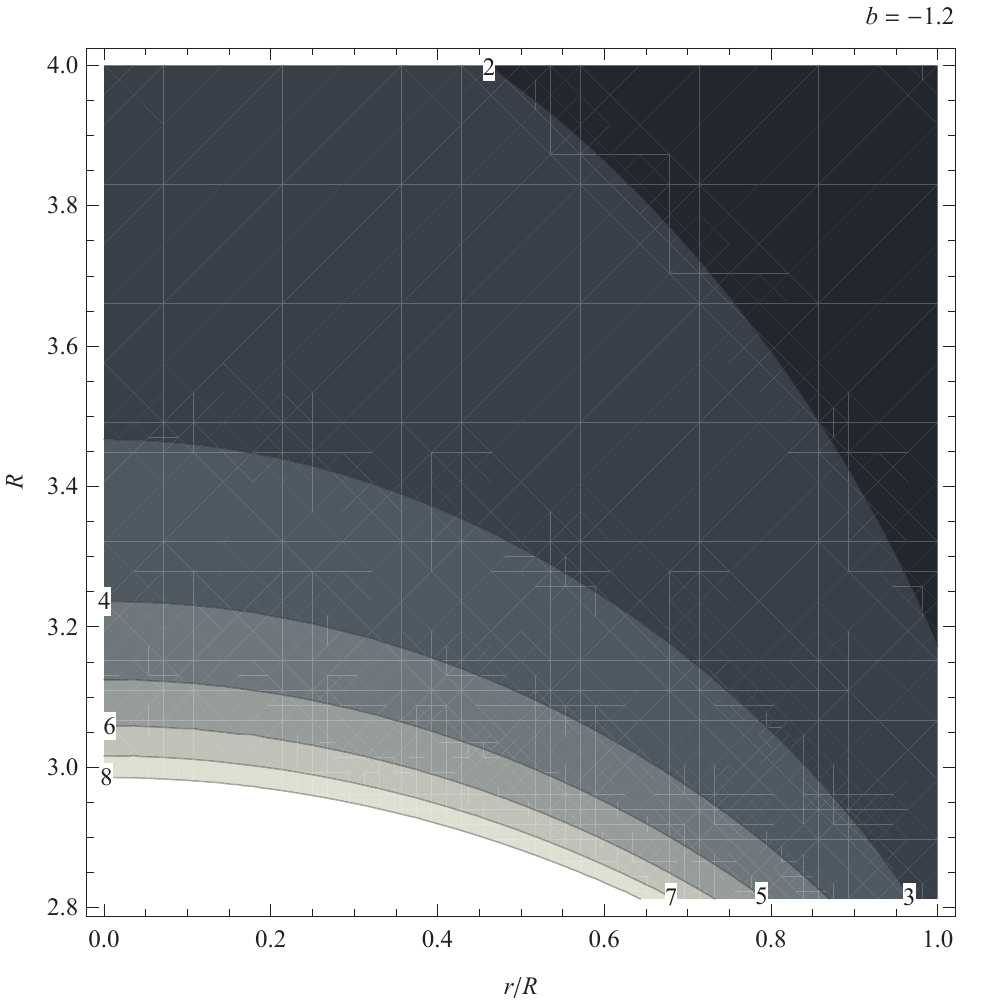}
        \centering\includegraphics[width=\hsize,keepaspectratio=true]{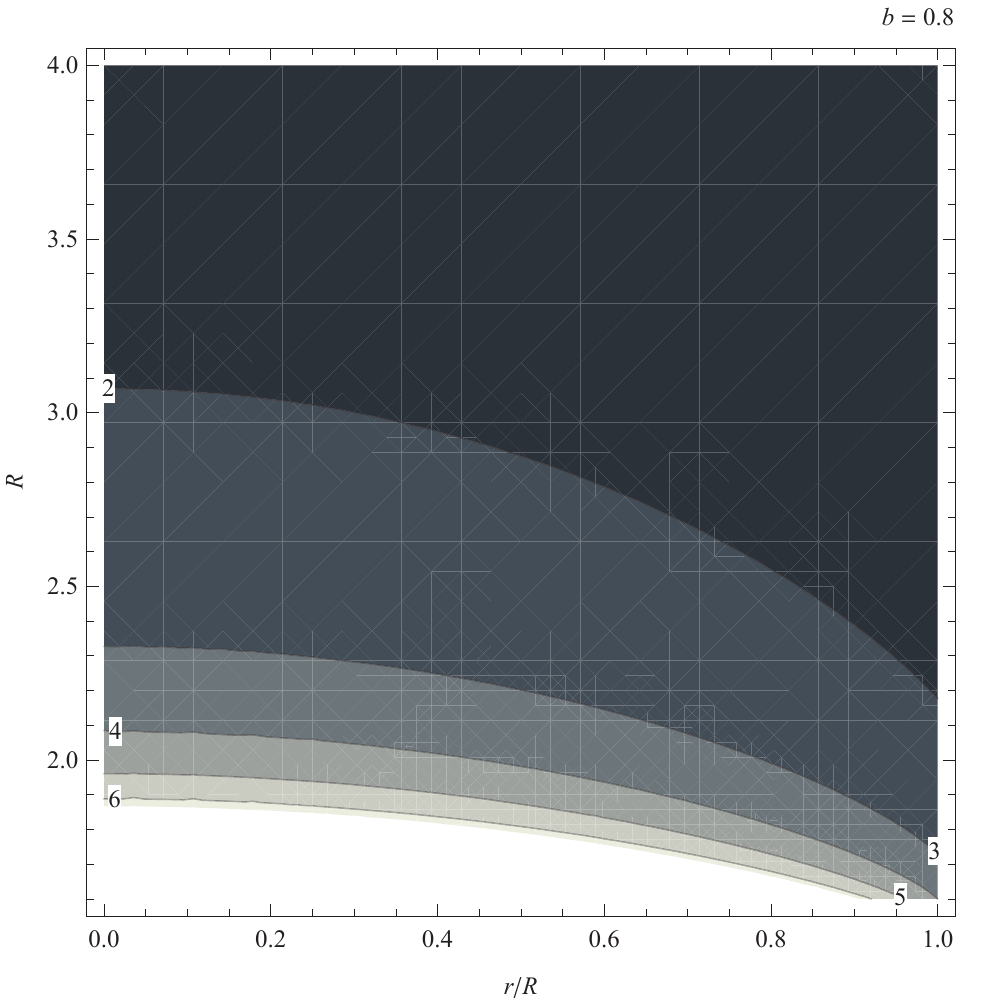}
      \end{minipage}\hfill%
      \begin{minipage}[b]{.445\hsize}
        \centering\includegraphics[width=\hsize,keepaspectratio=true]{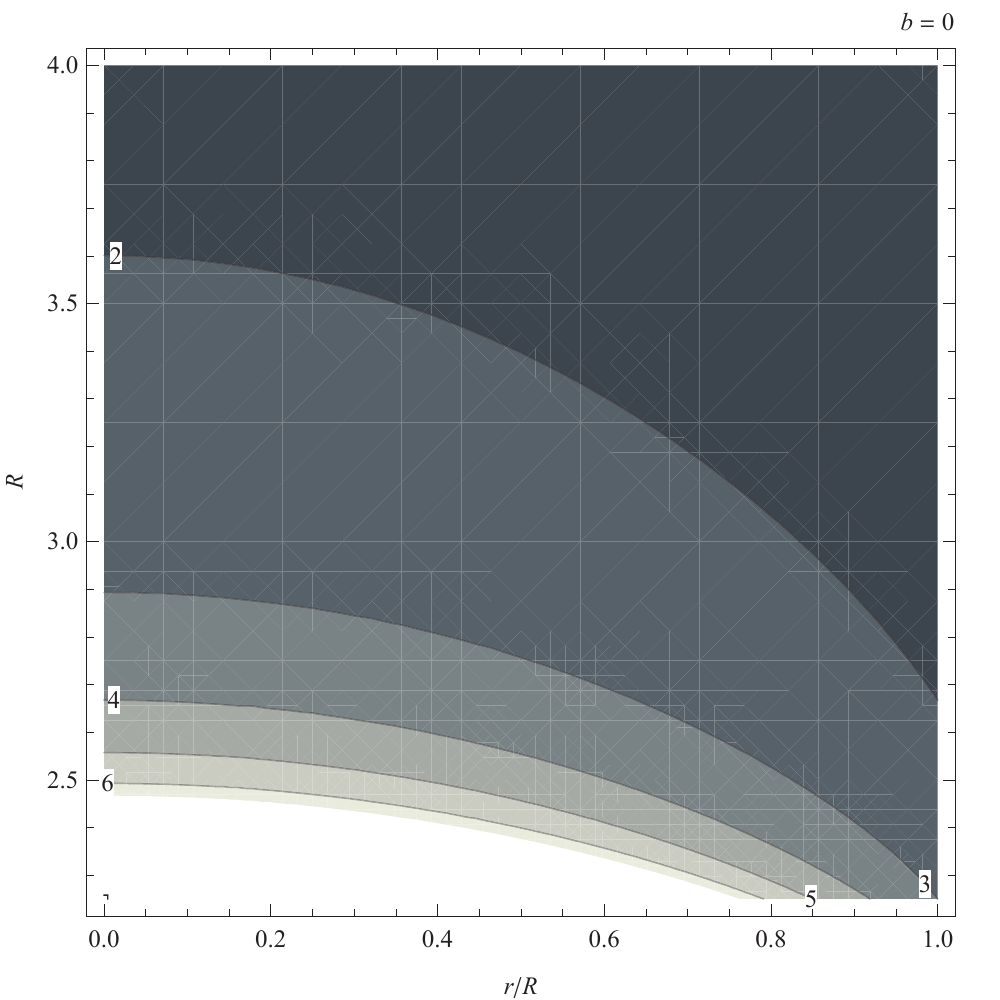}
        \centering\includegraphics[width=\hsize,keepaspectratio=true]{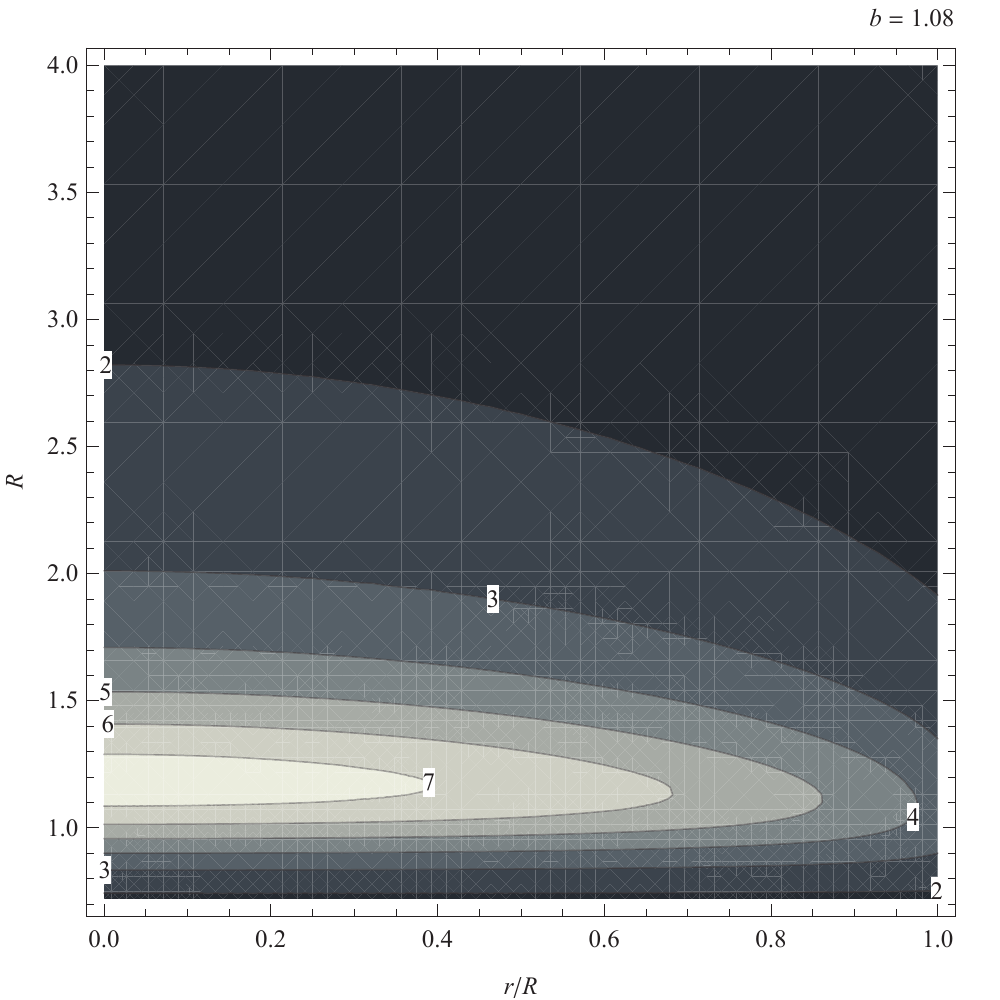}
      \end{minipage}
      \centering\includegraphics[width=.445\hsize,keepaspectratio=true]{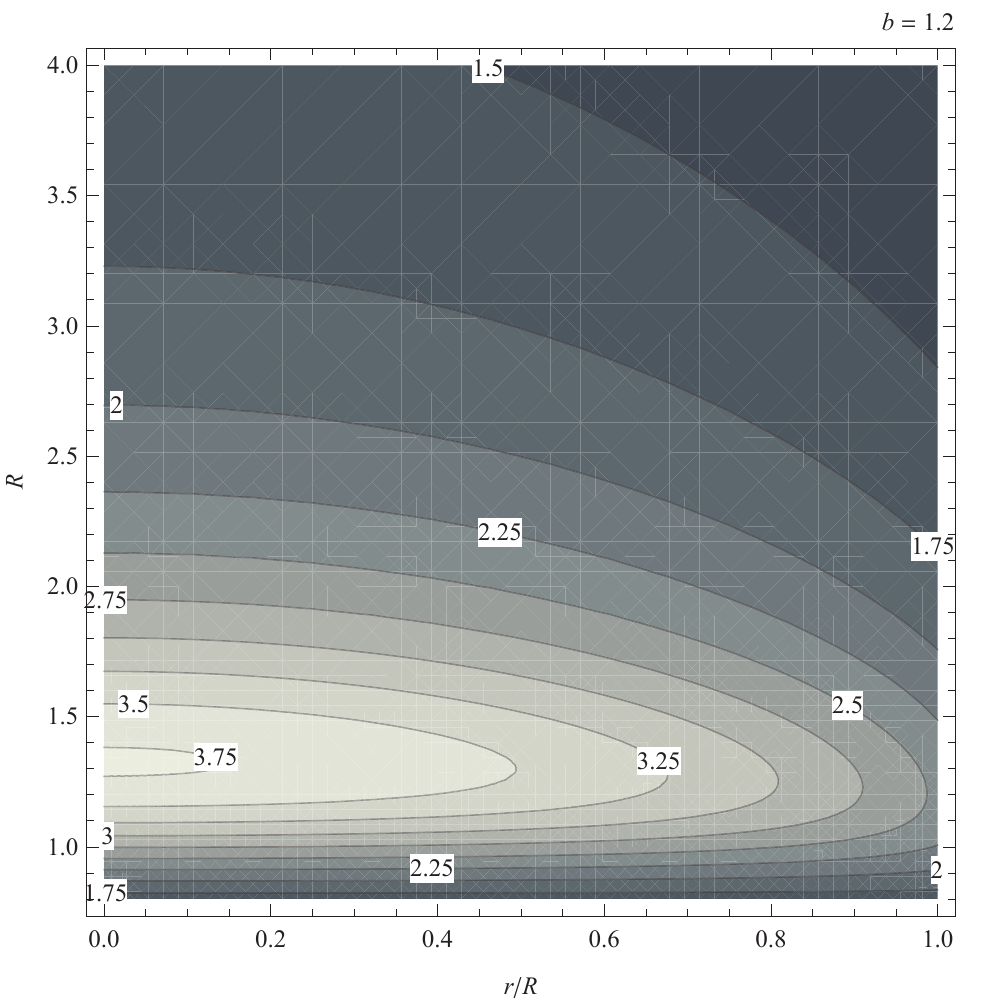}
      \caption{Contour plots of the redshift profiles $(1 + z)(r/R, R, b)$ for selected representative values of $b$. The area where surface extend beyond the plotting range is colored white. Notice the qualitative different character of the redshift profiles for compact stars with $b>1$.}\label{figure7}
    \end{figure}

    Because for given $b$, $R$ the pressure function $p(r)$ is monotonously decreasing function of increasing $r$, the values of redshifts are varying between those two limiting values. Behaviour of the central pressure is demonstrated in Figure~\ref{figure8}. In the case of compact stars with external spacetime of the black-hole type ($b < 1$), the central redshift diverges for $R \rightarrow R_\mathrm{min}(b)$. On the other hand, in the case of external spacetime of the naked-singularity type ($b > 1$), the central pressure remains finite for $R \rightarrow R_\mathrm{min}(b)$; in fact it reaches a maximum value for $R > R_\mathrm{min}(b)$. This special behaviour is also reflected by redshift profile given for fixed representative values of $b$ in Figure~\ref{figure7}.

\afterpage{\clearpage}

    It is convenient to find the redshift formula for compact stars with some special values of the surface radius related to the existence of extremely compact stars (when special phenomenon of neutrino trapping inside the star is important~\cite{Stu-Hla-Urb:2011:} and the photon circular orbits that can be relevant for accretion and optical phenomena outside the compact star).

    The central redshifts when the surface radius $R$ is located at the photon circular orbits of the naked-singularity spacetime (with $b > 1$) takes the form
    \begin{equation}
      (1 + z)(r = 0, R = r_\mathrm{ph+}, b) = \frac{\left(-4 b + \sqrt{9 - 8 b} + 3\right) \left(3 \sqrt{2} \sqrt{\frac{4 b + \sqrt{9 - 8 b} - 3}{b}} + 4\right)}{24 (1 - b)},
    \end{equation}

    \begin{equation}
      (1 + z)(r = 0, R = r_\mathrm{ph-}, b) = \frac{4 \sqrt{2} \left(\sqrt{2} \sqrt{\frac{4 b - \sqrt{9 - 8 b} - 3}{b}} b - 8 b - 4 \sqrt{9 - 8 b} + 12\right)}{3 \left(\sqrt{2} \sqrt{\frac{4 b - \sqrt{9 - 8 b} - 3}{b}} - 4\right) \sqrt{\frac{4 b - \sqrt{9 - 8 b} - 3}{b}} \left(2 b + \sqrt{9 - 8 b} - 3\right)}.
    \end{equation}

    For the limiting, maximally (minimally) extended extremely compact braneworld spacetimes the value of the central and minimal (maximal) redshift is given by $(1 + z)_\mathrm{min}^{\mathrm{ECS}}$ ($(1 + z)_\mathrm{max}^{\mathrm{ECS}}$) that are determined by
    \begin{equation}
      (1 + z){}_\mathrm{min}^{\mathrm{ECS}} =\left\{
        \begin{array}{lll}
          (1 + z)(r =0, R = r_\mathrm{ph+}, b) & \quad\mbox{for} & b \leq 0,\\
          (1 + z)(r =0, R = R_\mathrm{I+}, b) & \quad\mbox{for} & 0 < b < 27/20,
        \end{array}
      \right.
    \end{equation}
    where
    \begin{equation}
      (1 + z)(r = 0, R = R_\mathrm{I+}, b) = \frac{-\sqrt{6} \left(-4 b + \sqrt{81 - 60 b} + 9\right) - \frac{\left(\sqrt{81 - 60 b} + 9\right) b}{\sqrt{-4 b + \sqrt{81 - 60 b} + 9}}}{\sqrt{\frac{3}{2}} \left(-6 b + \sqrt{81 - 60 b} + 9\right) - \frac{9 \left(-4 b + \sqrt{81 - 60 b} + 9\right)^{3/2}}{\sqrt{81 - 60
       b} + 9}},
    \end{equation}
    and by
    \begin{equation}
      (1 + z){}_\mathrm{max}^{\mathrm{ECS}} =\left\{
        \begin{array}{lll}
          (1 + z)(r =0, R = R_\mathrm{min}, b) \rightarrow \infty & \quad\mbox{for} & b \leq 0,\\
          (1 + z)(r =0, R = R_\mathrm{I-}, b) & \quad\mbox{for} & 0 < b < 27/20,
        \end{array}
      \right.
    \end{equation}
    where
    \begin{equation}
        (1 + z)(r = 0, R = R_\mathrm{I-}, b) = \frac{\sqrt{6} \left(4 b + \sqrt{81 - 60 b} - 9\right) - \sqrt{\frac{\left(\sqrt{81 - 60 b} - 9\right)^2}{-4 b - \sqrt{81 - 60 b} + 9}}   b}{\sqrt{\frac{3}{2}} \left(-6 b - \sqrt{81 - 60 b} + 9\right) - 9 \sqrt{\frac{\left(-4 b - \sqrt{81 - 60 b} + 9\right)^3}{\left(\sqrt{81 - 60 b} - 9\right)^2}}}.
    \end{equation}

    We give the central redshift limiting values in dependence on the braneworld parameter $b$ in Figure~\ref{figure5}. The central redshift of compact stars with surface located at the photon circular orbits of the external spacetime is illustrated in dependence on the tidal charge in Figure~\ref{figure6}. Notice that for compact stars with $b \in (1, 1.0749)$ the surface redshift from the compact star with $R = r_\mathrm{ph-}$ overcomes the central redshift of the compact star with $R = r_\mathrm{ph+}$.

    \section{Discussion}\label{SECconcl}

    \begin{figure}[t]%
      \centering\includegraphics[width=0.499\hsize]{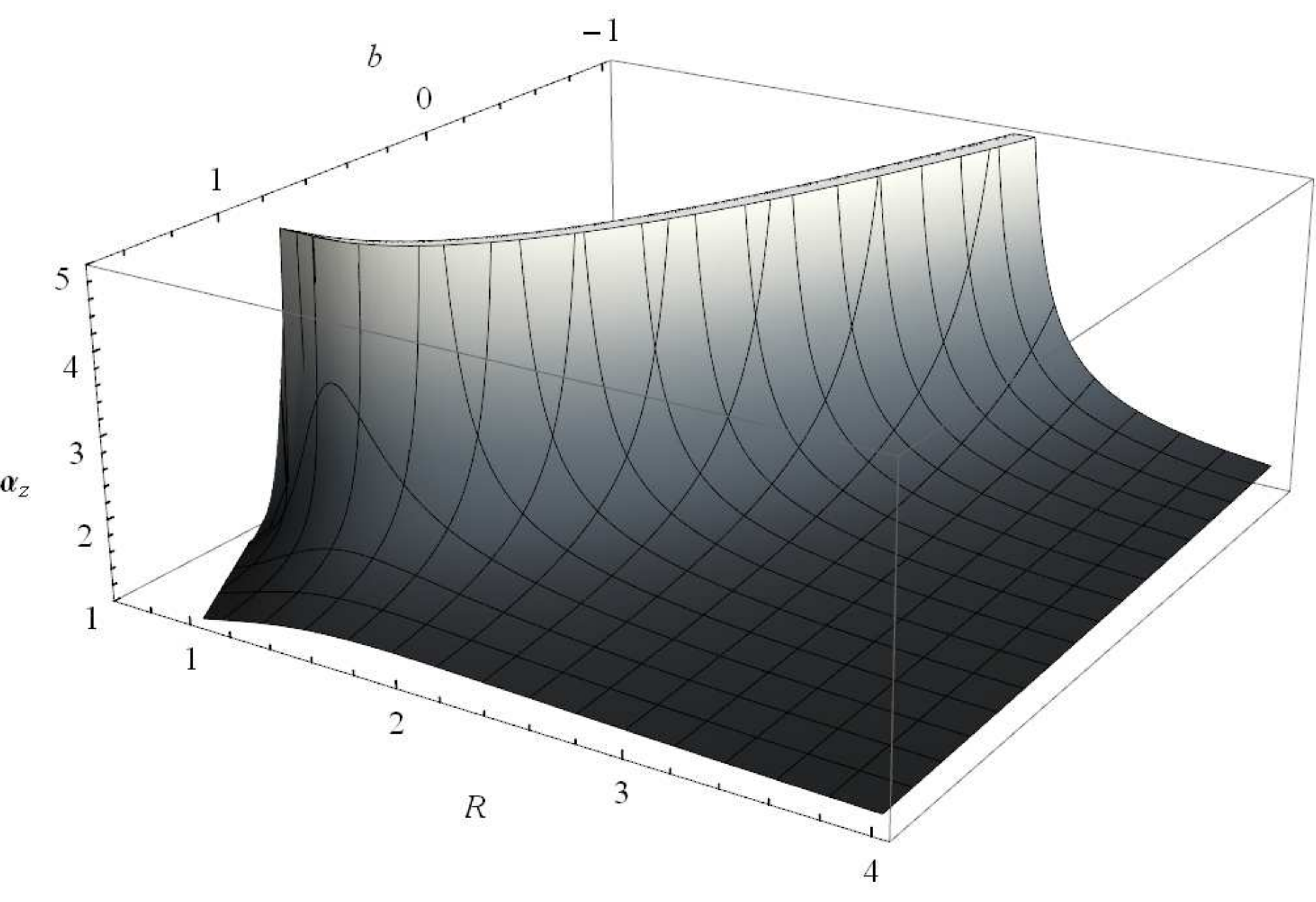}\includegraphics[width=0.499\hsize]{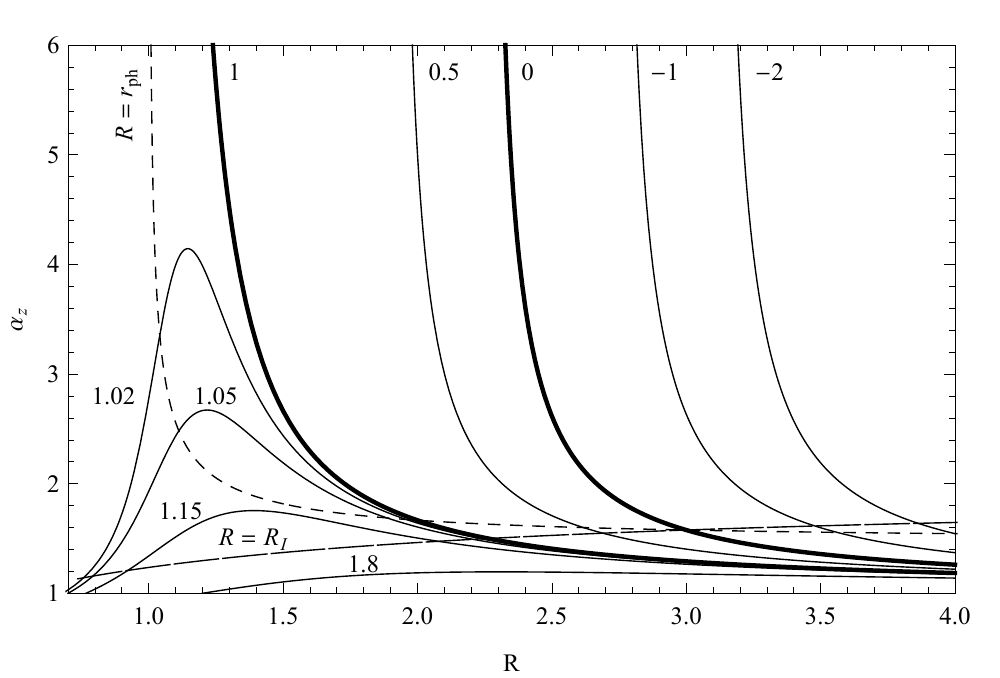}
      \caption{The parameter $\alpha_\mathrm{z}$ relating the central and surface redshift given as a function of tidal charge $b$ and surface of radius $R$. On the plot on the right the values of parameter $b$ are given near the correspondig curves.}\label{figure9}
    \end{figure}

    \subsection{The role of the tidal charge}
    In order to obtain a theoretical tool for estimating possible presence of the tidal charge influence using observational data on the surface redshift of photons and internal redshift of neutrinos, it could be useful to compare the surface redshift $(1 + z)(r = R, R, b)$ to the central redshift $(1 + z)(r = 0, R, b)$ and the internal redshift interval
    \begin{equation}
      \Delta_z (R, b) \equiv (1 + z)(r = 0, R, b) - (1 + z)(r = R, R, b) = \frac{p(r = 0, R, b)}{\varrho\Delta(R, b)}.
    \end{equation}
    We can introduce a characteristic quantity
    \begin{equation}
      \alpha_\mathrm{z}(R, b) \equiv \frac{(1 + z)(r = 0, R, b)}{(1 + z)(r = R, R, b)}= 1 + \frac{p(r = 0, R, b)}{\varrho}.
    \end{equation}

    The plot of $\alpha_\mathrm{z}(R, b)$ is given in Figure~\ref{figure9}. We see that in ECS with $R \rightarrow R_\mathrm{min}(b\leq1)$ the parameter $\alpha$ diverges while in compact stars with the naked-singularity type external spacetime the redshift remains finite for whole range of allowed surface radii. The plot of $\alpha_z$ is given in Figure~\ref{figure9} also explicitly for $R = r_\mathrm{ph}$ and $R = R{}_\mathrm{I}$.

    \begin{figure}[t]%
      \centering\includegraphics[width=.6\hsize]{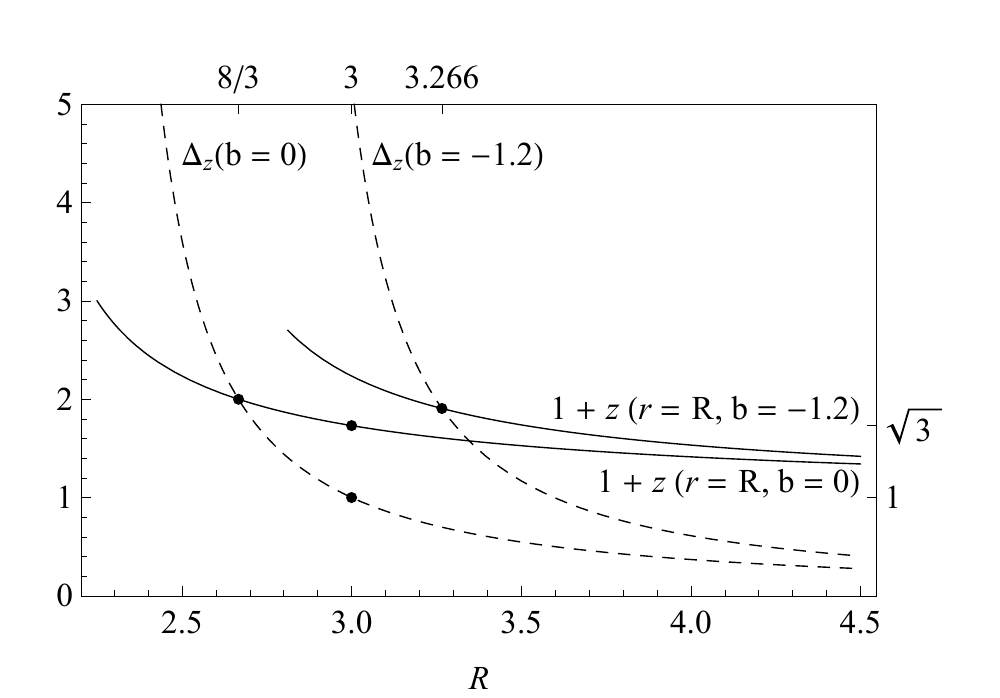}\\
      \centering\includegraphics[width=.499\hsize]{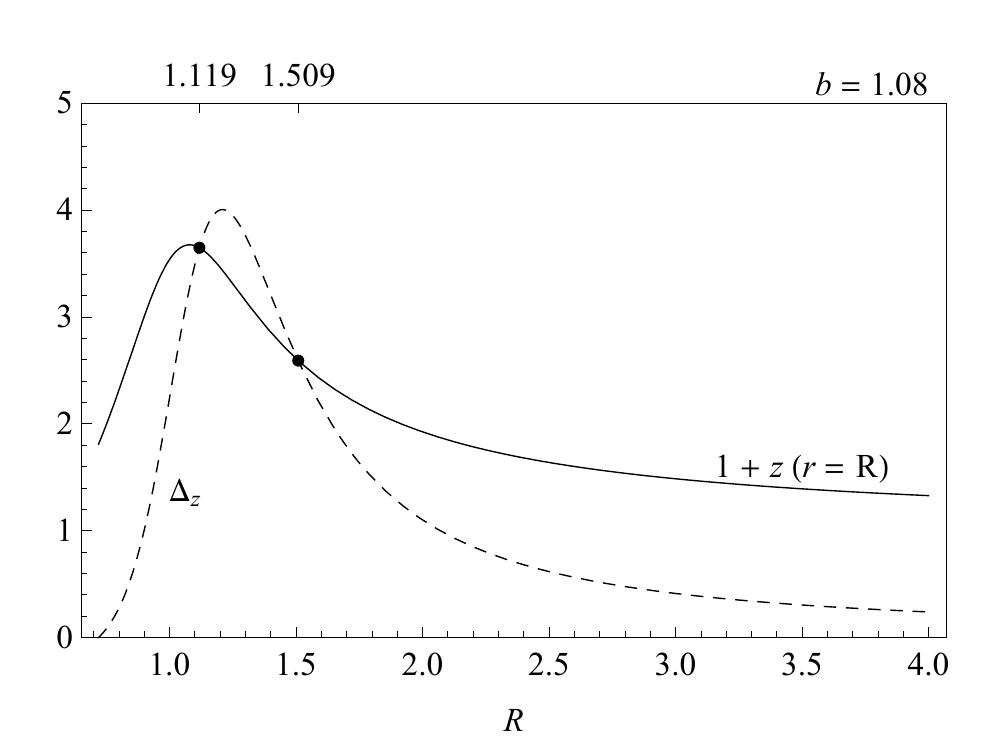}\includegraphics[width=.499\hsize]{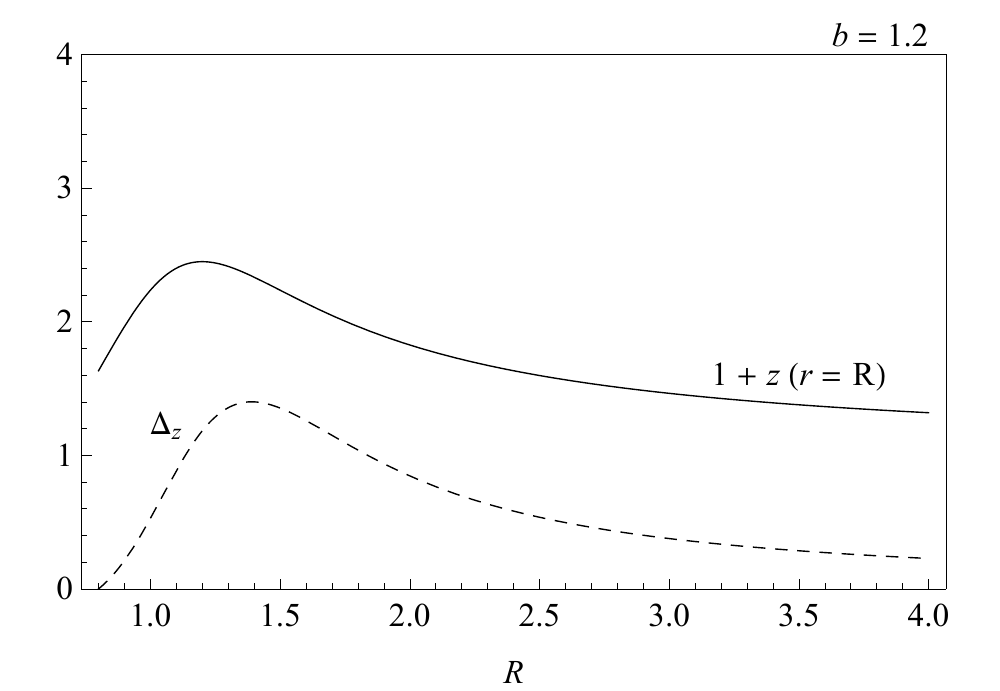}
      \caption{Plot of the surface redshifts $1 + z$ and the difference of central and surface redshift $\Delta_\mathrm{z}$ for $b = -1.2, 0, 1.08, 1.2$.}\label{figure10}
    \end{figure}

    It is useful to relate the surface redshift and the redshift difference $\Delta_\mathrm{z}$. In the standard Schwarzschild spacetimes ($b = 0$), the $(1 + z)(r = R, R, b = 0)$ and $\Delta_\mathrm{z}(R, b = 0)$ are illustrated in Figure~\ref{figure10}. We can see that for the maximal limit of ECS ($R = r_\mathrm{ph}$), there is $1 + z = \sqrt{3}$ and $\Delta_\mathrm{z} = 1$. For $R > r_\mathrm{ph}$ increasing $(1 + z)(r = R, R, b = 0)$ decreases slowly, but $\Delta_\mathrm{z}$ decreases fast to zero. On the other hand, for $R \rightarrow R_\mathrm{min} = 9/4$, the surface redshift slowly increases to $(1 + z)(r = R\rightarrow R_\mathrm{min}) = 3$, but $\Delta_\mathrm{z}$ fast diverges. There is $(1+z)(r = R, R, b)=\Delta_\mathrm{z}(R)$ for $R = 8/3$. For negative (positive) tidal charges both the surface redshift (1 + z)(r = R) and the difference $\Delta_\mathrm{z}(R, b)$ decrease (increases) with decreasing (increasing) $b$. Notice that for $b \in (1, 1.10254)$ there are two common points of $(1 + z)(r = R)$ and $\Delta_\mathrm{z}$, while for $b > 1.10254$ there are no common points.

  \subsection{Determination of the compact star parameters}
    \begin{figure}[t]%
      \centering\includegraphics[width=.7\hsize]{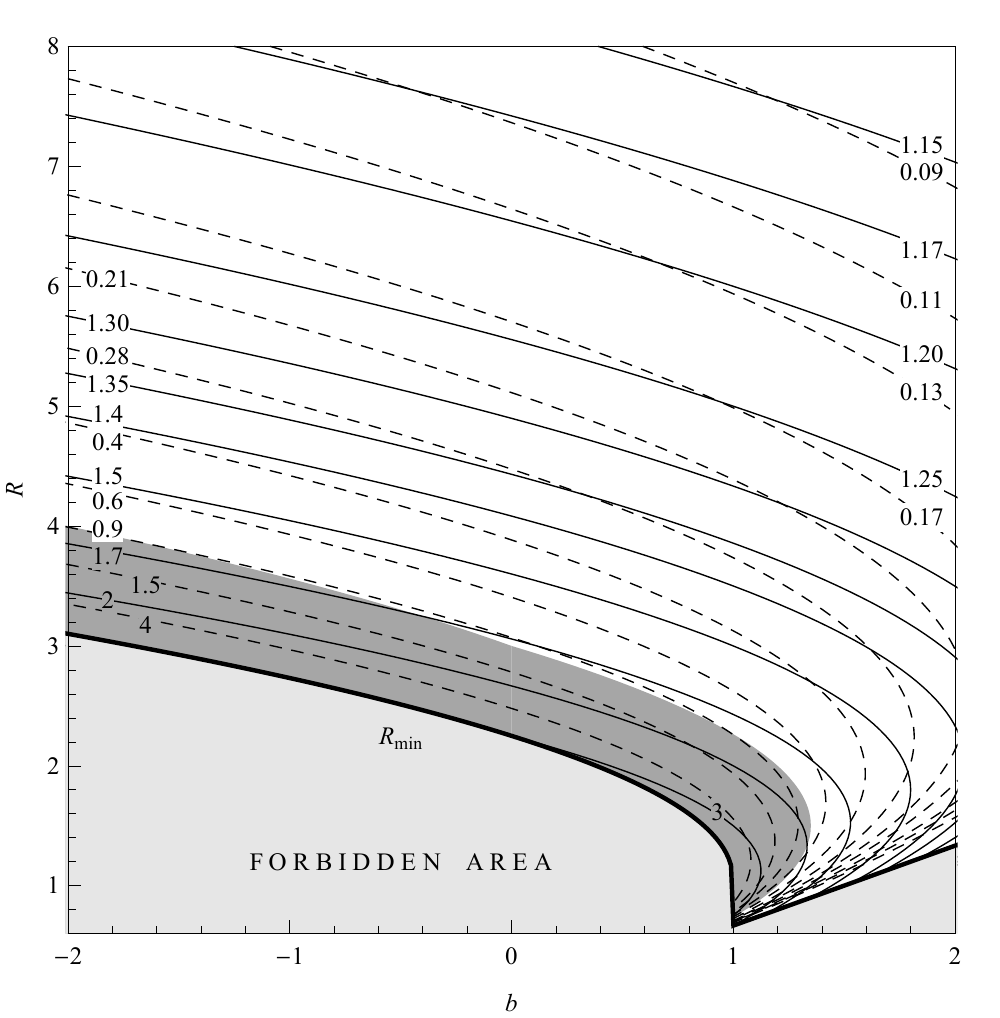}
      \caption{Plot for graphical determination of parameters $R$ and $b$ from values $\Delta_z$ and $(1 + z)(r = R, R, b)$. Area of ECS is denoted by dark gray colour.}\label{figure11}
    \end{figure}
    From the knowledge of values $\Delta_z$ and $(1 + z)(r = R, R, b)$ (both of them are in principle observable measurable) we can determine the radius of the star $R$ and braneworld parameter $b$ via relations ($(1 + z)_R \equiv (1 + z)(r = R, R, b)$)
    \begin{equation}
      R =  \frac{(1 + z)_{R}^{2} \left[3 \Delta_z + (1 + z)_R\right]}{\left[(1 + z)_R + 1\right] \left\{3 \Delta_z \left[(1 + z)_R - 2\right] + 2 \left[(1 + z)_R - 1\right](1 + z)_R\right\}},
    \end{equation}
    and
    \begin{equation}
      b =  \frac{3 (1 + z)_R^2 \left[3 \Delta_z + (1 + z)_R\right] \left\{\Delta_z \left[(1 + z)_R - 3\right] + \left[(1 + z)_R - 1\right] (1 + z)_R\right\}}{\left[(1 + z)_R + 1\right]\left\{3 \Delta_z \left[(1 + z)_R - 2\right] + 2 \left[(1 + z)_R - 1\right] (1 + z)_R\right\}^2}.
    \end{equation}
    The results obtained by this procedure can be confronted with relations derived above for limiting extremely compact stars and the compact star with surface coinciding with the photon circular orbits of the external spacetime. Determination of the parameters $R$ and $b$ from the observationally determined quantities $\Delta_z$ and $(1 + z)(r = R, R, b)$ can be represented by nomograms. We illustrate the nomogram in Figure~\ref{figure11} where the region of ECS is explicitly depicted and we can immediately find if the observed compact star could belong to the ECS, or could allow for an unusual character of accretion phenomena in vicinity of compact stars having their surface located under the photon circular orbits of the external spacetime.

  \section{Conclusions}
    In our study we have focused our attention on the simple model of the braneworld neutron (quark) stars and we have studied the surface (photon) redshift and the internal (neutrino) redshift under assumption of zero neutrino rest energy and the motion along null geodesics. We postpone to future investigations the realistic case of non-zero neutrino rest energy and the role of neutrino flavour mixing along the trajectory to the observers \cite{alt:2007:,nuno:2007:,thomas:2008:}. (In any case it is reasonable to neglect the effect of neutrino flavour missing on the short distances inside the compact stars.) We believe that our results will keep relevance even in such more complex realistic situations. Of special interest could be a study of the redshift phenomena related to the full 5D uniform energy density stars numerically constructed in~\cite{Wis:2002:}.

    We can conclude that our results indicate at least a principal possibility to find the compact star parameters (both surface radius and the tidal charge) from measurements of photon surface redshift and the range of internal redshift of neutrinos. From the observational data we are able to decide if the observed braneworld compact star is an extremely compact star allowing for gravitational trapping of some part of radiated neutrinos. Of course, the results giving the compact stars parameters from internal phenomena have to be further confronted to the observational phenomena related to accretion discs orbiting the compact stars where the tidal charge can yield some other specific signatures related to optical effects of accretion discs.

\clearpage

  \acknowledgments{The present work was supported by the Czech grants MSM~4781305903, LC~06014, GA\v{C}R 205/09/H033 and the internal grant SGS/2/2010. One of the authors (Z.\,S.) would like to express his gratitude to the Czech Committee for collaboration with CERN and the Theory Division of CERN for perfect hospitality.}

\bibliography{references}

\providecommand{\href}[2]{#2}\begingroup\raggedright\begin{thebibliography}{10}

\bibitem{Ark-Dim-Dva:1998:}
N.~{Arkani-Hamed}, S.~{Dimopoulos}, and G.~{Dvali}, {\it {The hierarchy problem
  and new dimensions at a millimeter}},  {\em Phys. Lett. \emph{B}} {\bf 429}
  (1998) 263--272.

\bibitem{Ran-Sun:1999:}
L.~{Randall} and R.~{Sundrum}, {\it {An Alternative to Compactification}},
  {\em Phys. Rev. Lett.} {\bf 83} (1999) 4690--4693.

\bibitem{Shi-Mae-Sas:1999:}
T.~{Shiromizu}, K.-i. {Maeda}, and M.~{Sasaki}, {\it {The Einstein equations on
  the 3-brane world}},  {\em Phys. Rev. \emph{D}} {\bf 62} (2000) 024012.

\bibitem{Maar:2004:}
R.~{Maartens}, {\it {Brane-World Gravity}},  {\em Living Rev. Relativity} {\bf
  7} (2004) 7.

\bibitem{Dad-Maar-Pap-Rez:2000:PHYSR4:}
N.~{Dadhich}, R.~{Maartens}, P.~{Papadopoulos}, and V.~{Rezania}, {\it {Black
  holes on the brane}},  {\em Phys. Lett. \emph{B}} {\bf 487} (2000) 1--6.

\bibitem{Ali-Gum:2005:}
A.~N. {Aliev} and A.~E. {G{\"u}mr{\"u}k{\c c}{\"u}o{\u g}lu}, {\it {Charged
  rotating black holes on a 3-brane}},  {\em Phys. Rev. \emph{D}} {\bf 71}
  (2005) 104027.

\bibitem{Ger-Maar:2001:}
C.~{Germani} and R.~{Maartens}, {\it {Stars in the braneworld}},  {\em Phys.
  Rev. \emph{D}} {\bf 64} (2001) 124010.

\bibitem{Stu-Kot:2009:}
Z.~{Stuchl{\'{\i}}k} and A.~{Kotrlov{\'a}}, {\it {Orbital resonances in discs
  around braneworld Kerr black holes}},  {\em Gen. Relativ. Gravit.} {\bf 41}
  (2009) 1305--1343.

\bibitem{Ali-Tal:2009:}
A.~N. {Aliev} and P.~{Talazan}, {\it {Gravitational effects of rotating
  braneworld black holes}},  {\em Phys. Rev. \emph{D}} {\bf 80} (2009) 044023.

\bibitem{Abdu-Ahme:2010:PHYSR4}
A.~{Abdujabbarov} and B.~{Ahmedov}, {\it {Test particle motion around a black
  hole in a braneworld}},  {\em Phys. Rev. \emph{D}} {\bf 81} (2010) 044022.

\bibitem{Mam-Hak-Toj:2010:MPLA:}
A.~I. {Mamadjanov}, A.~A. {Hakimov}, and S.~R. {Tojiev}, {\it {Quantum
  Interference Effects in Spacetime of Slowly Rotating Compact Objects in
  Braneworld}},  {\em Mod. Phys. Lett. \emph{A}} {\bf 25} (2010) 243--256.

\bibitem{Mor-Ahme-Abdu-Mam:2010:ASS:}
V.~S. {Morozova}, B.~J. {Ahmedov}, A.~A. {Abdujabbarov}, and A.~I.
  {Mamadjanov}, {\it {Plasma magnetosphere of rotating magnetized neutron star
  in the braneworld}},  {\em Astropys. \&{} Space Sci.} {\bf 330} (2010)
  257--266.

\bibitem{Sche-Stu:2009:a}
J.~{Schee} and Z.~{Stuchl{\'{\i}}k}, {\it {Optical Phenomena in the Field of
  Braneworld Kerr Black Holes}},  {\em Int. J. Mod. Phys. \emph{D}} {\bf 18}
  (2009) 983--1024.

\bibitem{Sche-Stu:2009:b}
J.~{Schee} and Z.~{Stuchl{\'{\i}}k}, {\it {Profiles of emission lines generated
  by rings orbiting braneworld Kerr black holes}},  {\em Gen. Relativ. Gravit.}
  {\bf 41} (2009) 1795--1818.

\bibitem{Bin-Nun:2010:PHYSR4:a}
A.~Y. {Bin-Nun}, {\it {Gravitational lensing of stars orbiting Sgr A${}^*$ as a
  probe of the black hole metric in the Galactic center}},  {\em Phys. Rev.
  \emph{D}} {\bf 82} (2010) 064009.

\bibitem{Bin-Nun:2010:PHYSR4:b}
A.~Y. {Bin-Nun}, {\it {Relativistic images in Randall-Sundrum II braneworld
  lensing}},  {\em Phys. Rev. \emph{D}} {\bf 81} (2010) 123011.

\bibitem{Wis:2002:}
T.~{Wiseman}, {\it {Relativistic stars in Randall-Sundrum gravity}},  {\em
  Phys. Rev. \emph{D}} {\bf 65} (2002) 124007.

\bibitem{Wis:2003:}
T.~{Wiseman}, {\it {Static axisymmetric vacuum solutions and non-uniform black
  strings}},  {\em Class. Quant. Grav.} {\bf 20} (2003) 1137--1175.

\bibitem{Kud-Tan-Nak:2003:}
H.~{Kudoh}, T.~{Tanaka}, and T.~{Nakamura}, {\it {Small localized black holes
  in a braneworld: Formulation and numerical method}},  {\em Phys. Rev.
  \emph{D}} {\bf 68} (2003) 024035.

\bibitem{Kud:2004:}
H.~{Kudoh}, {\it {Six-dimensional localized black holes: Numerical solutions}},
   {\em Phys. Rev. \emph{D}} {\bf 69} (2004) 104019.

\bibitem{Fig-Wis:2011:}
P.~{Figueras} and T.~{Wiseman}, {\it {Gravity and large black holes in
  Randall-Sundrum II braneworlds}},  {\em ArXiv e-prints} (May, 2011)
  [\href{http://xxx.lanl.gov/abs/1105.2558}{{\tt arXiv:1105.2558}}].

\bibitem{Kot-Stu-Tor:2008:CLASQG:}
A.~{Kotrlov{\'a}}, Z.~{Stuchl{\'{\i}}k}, and G.~{T{\"o}r{\"o}k}, {\it
  {Quasiperiodic oscillations in a strong gravitational field around neutron
  stars testing braneworld models}},  {\em Class. Quantum Grav.} {\bf 25}
  (2008) 225016.

\bibitem{Boh-Har-Lob:2008:CLAQG:}
C.~G. {B{\"o}hmer}, T.~{Harko}, and F.~S.~N. {Lobo}, {\it {Solar system tests
  of brane world models}},  {\em Class. Quantum Grav.} {\bf 25} (2008) 045015.

\bibitem{Boh-Ris-Har-Lob:2010:CLASQG}
C.~G. {B{\"o}hmer}, G.~{De Risi}, T.~{Harko}, and F.~S.~N. {Lobo}, {\it
  {Classical tests of general relativity in brane world models}},  {\em Class.
  Quantum Grav.} {\bf 27} (2010) 185013.

\bibitem{Stu-Hla-Urb:2011:}
Z.~{Stuchl{\'{\i}}k}, J.~{Hlad{\'{\i}}k}, and M.~{Urbanec}, {\it {Neutrino
  trapping in braneworld extremely compact stars}},  {\em Gen. Relativ.
  Gravit.} (2011). Submitted.

\bibitem{Lat-Pra:2007:PhysRep:}
J.~M. {Lattimer} and M.~{Prakash}, {\it {Neutron star observations: Prognosis
  for equation of state constraints}},  {\em Phys. Rep.} {\bf 442} (2007)
  109--165.

\bibitem{Abr-Mil-Stu:1993:PHYSR4:}
M.~A. {Abramowicz}, J.~C. {Miller}, and Z.~{Stuchl{\'{\i}}k}, {\it {Concept of
  radius of gyration in general relativity}},  {\em Phys. Rev. \emph{D}} {\bf
  47} (1993) 1440--1447.

\bibitem{Stu:2000:ACTPS2:}
Z.~{Stuchl{\'{\i}}k}, {\it {Spherically Symmetric Static Configurations of
  Uniform Density in Spacetimes with a Non-Zero Cosmological Constant}},  {\em
  Acta Phys. Slov.} {\bf 50} (2000) 219--228.

\bibitem{Nil-Cla:2000:GRRelStarsPolyEOS:}
U.~S. {Nilsson} and C.~{Uggla}, {\it {General Relativistic Stars: Polytropic
  Equations of State}},  {\em Annals of Physics} {\bf 286} (2000) 292--319.

\bibitem{Stu-Tor-Hle-Urb:2009:}
Z.~{Stuchl{\'{\i}}k}, G.~{T{\"o}r{\"o}k}, S.~{Hled{\'{\i}}k}, and M.~{Urbanec},
  {\it {Neutrino trapping in extremely compact objects: I. Efficiency of
  trapping in the internal Schwarzschild spacetimes}},  {\em Class. Quantum
  Grav.} {\bf 26} (2009) 035003.

\bibitem{Bal-Bic-Stu:1989:}
V.~{Balek}, J.~{Bi\v{c}\'{a}k}, and Z.~{Stuchl\'{\i}k}, {\it {The motion of the
  charged particles in the field of rotating charged black holes and naked
  singularities. II - The motion in the equatorial plane}},  {\em Bull. Astron.
  Inst. Czechosl.} {\bf 40} (1989), no.~3 133--165.

\bibitem{Stu-Hle:2002:}
Z.~{Stuchl{\'{\i}}k} and S.~{Hled{\'{\i}}k}, {\it {Properties of the
  Reissner-Nordstr\"{o}m spacetimes with a nonzero cosmological constant}},
  {\em Acta Phys. Slov.} {\bf 52} (2002), no.~5 363--407.

\bibitem{Sha-Teu:1983:BHWDNS:}
S.~L. {Shapiro} and S.~A. {Teukolsky}, {\em {Black holes, white dwarfs, and
  neutron stars: The physics of compact objects}}.
\newblock Wiley--VCH, New York, 1983.

\bibitem{Gle:2000:CompactStars:}
N.~K. {Glendenning}, {\em {Compact stars : nuclear physics, particle physics,
  and general relativity}}.
\newblock Springer, New York, 2000.

\bibitem{Web:1999:Pul:}
F.~{Weber}, {\em {Pulsars as astrophysical laboratories for nuclear and
  particle physics}}.
\newblock Institute of Physics Pub., Bristol, U.\,K., 1999.

\bibitem{Stu-Schee:2010:CLAQG:}
Z.~{Stuchl{\'{\i}}k} and J.~{Schee}, {\it {Appearance of Keplerian discs
  orbiting Kerr superspinars}},  {\em Class. Quantum Grav.} {\bf 27} (2010)
  215017.

\bibitem{Stu-Hle:2000:}
Z.~{Stuchl{\'{\i}}k} and S.~{Hled{\'{\i}}k}, {\it {Equatorial photon motion in
  the Kerr-Newman spacetimes with a non-zero cosmological constant}},  {\em
  Class. Quantum Grav.} {\bf 17} (2000) 4541--4576.

\bibitem{alt:2007:}
G.~{Altarelli}, {\it {Lectures on Models of Neutrino Masses and Mixings}},
  {\em ArXiv e-prints} (2007) [\href{http://xxx.lanl.gov/abs/0711.0161}{{\tt
  arXiv:0711.0161}}].

\bibitem{nuno:2007:}
H.~{Nunokawa}, S.~{Parke}, and J.~W.~F. {Valle}, {\it {CP violation and
  neutrino oscillations}},  {\em Prog. Part. Nucl. Phys.} {\bf 60} (2008)
  338--402.

\bibitem{thomas:2008:}
P.~L. {Thomas} and J.~A. {Vahle}, {\em {Neutrino Oscillations: Present Status
  and Future Plans}}.
\newblock World Scientific Publishing, Singapore, 2008.

\end{thebibliography}\endgroup
\bibliographystyle{JHEP}
\end{document}